\documentclass[10pt]{scrartcl}
\usepackage[a4paper, total={16cm, 24cm}]{geometry}
\usepackage{amsmath,amssymb,amsthm,natbib,microtype}

\usepackage{booktabs} \usepackage{tikz}

\usepackage[pagebackref]{hyperref}
\usepackage[nameinlink]{cleveref}
\hypersetup{
	breaklinks=true,
	colorlinks,
	citecolor=green!45!black,
	linkcolor=red!60!black,
	urlcolor=purple!80!black
}

\usepackage{multirow}

\usepackage{nicefrac}
\usepackage{bbold}

\newcommand{\id}{\mathbb{I}}
\DeclareMathOperator{\D}{D}
\newcommand{\utilityclass}{\mathcal{U}}
\DeclareMathOperator{\unitsum}{unit-sum}
\DeclareMathOperator{\unitrange}{unit-range}
\DeclareMathOperator{\approval}{approval}
\DeclareMathOperator{\balanced}{balanced}
\DeclareMathOperator{\all}{all}

\DeclareMathOperator{\PF}{PF}
\DeclareMathOperator{\SW}{UW}
\DeclareMathOperator{\NW}{NW}

\DeclareMathOperator{\harm}{hsc}
\newcommand{\HR}{f_{\textup{HR}}}
\newcommand{\SLR}{f_{\textup{SLR}}}
\newcommand{\SCR}{f_{\textup{SCR}}}
\newcommand{\cSCR}{f_{c\textup{-SCR}}}
\newcommand{\NashTwo}{f_{\textup{2Nash}}}
\newcommand{\PFTwo}{f_{\textup{2PF}}}
\DeclareMathOperator{\rank}{rank}

\newcommand{\bbN}{\mathbb{N}}
\newcommand{\bbR}{\mathbb{R}}
\renewcommand{\succeq}{\succcurlyeq}
\renewcommand{\le}{\leqslant}

\renewcommand{\ge}{\geqslant}

\renewcommand{\epsilon}{\varepsilon}

\newcommand{\set}[1]{\{#1\}}

\newcommand{\XX}{\mathbf{X}}

\newcommand{\xx}{\mathbf{x}}
\newcommand{\yy}{\mathbf{y}}

\newcommand{\ww}{\vec{w}}
\newcommand{\zz}{\mathbf{z}}

\newcommand{\vzz}{\vec{z}}

\newcommand{\voters}{N}
\newcommand{\voter}{i}
\newcommand{\candids}{A}

\newcommand{\power}{\mathcal{P}}

\newcommand{\utilities}{\vec{u}}
\newcommand{\induces}{\triangleright}

\newcommand{\pref}{\sigma}
\newcommand{\profile}{\vec{\sigma}}

\DeclareMathOperator{\E}{\mathbb{E}}

\newcommand{\UWTwoUnitsum}{f^{\unitsum}_{\textup{2-UW}}}
\newcommand{\UWTwoUnitrange}{f^{\unitrange}_{\textup{2-UW}}}

\DeclareMathOperator{\poly}{poly}
\DeclareMathOperator*{\argmax}{arg\,max}
\DeclareMathOperator*{\argmin}{arg\,min}

\DeclareMathOperator{\payoff}{R}

\theoremstyle{plain}
\newtheorem{theorem}{Theorem}[section]
\newtheorem{proposition}[theorem]{Proposition}
\newtheorem{lemma}[theorem]{Lemma}
\newtheorem{corollary}[theorem]{Corollary}
\theoremstyle{definition}
\newtheorem{definition}[theorem]{Definition}
\newtheorem{example}[theorem]{Example}

\title{Optimized Distortion and \\ Proportional Fairness in Voting}

\usepackage{authblk}

\author[1]{Soroush Ebadian}
\author[2]{Anson Kahng}
\author[3]{\authorcr Dominik Peters}
\author[1]{Nisarg Shah}
\affil[1]{University of Toronto} 
\affil[2]{University of Rochester} 
\affil[3]{CNRS, LAMSADE, Universit\'e Paris Dauphine - PSL}

\date{\vspace{-12pt}\sffamily \large Version v3 as accepted by ACM Transactions on Economics and Computation \\ \medskip January 2024\vspace{-16pt}}

\begin{document}

\maketitle

\begin{abstract}
	A voting rule decides on a probability distribution over a set of $m$ alternatives, based on rankings of those alternatives provided by agents. 
	We assume that agents have cardinal utility functions over the alternatives, but voting rules have access to only the rankings induced by these utilities.
	We evaluate how well voting rules do on measures of social welfare and of proportional fairness, computed based on the hidden utility functions.
	
	In particular, we study the \emph{distortion} of voting rules, which is a worst-case measure. It is an approximation ratio comparing the utilitarian social welfare of the optimum outcome to the social welfare produced by the outcome selected by the voting rule, in the worst case over possible input profiles and utility functions that are consistent with the input. The previous literature has studied distortion with unit-sum utility functions (which are normalized to sum to $1$), and left a small asymptotic gap in the best possible distortion. Using tools from the theory of fair multi-winner elections, we propose the first voting rule which achieves the optimal distortion $\Theta(\sqrt{m})$ for unit-sum utilities. Our voting rule also achieves optimum $\Theta(\sqrt{m})$ distortion for a larger class of utilities, including unit-range and approval (0/1) utilities.
	
	We then take a similar worst-case approach to a quantitative measure of the fairness of a voting rule, called \emph{proportional fairness}. Informally, it measures whether the influence of cohesive groups of agents on the voting outcome is proportional to the group size. We show that there is a voting rule which, without knowledge of the utilities, can achieve a $\Theta(\log m)$-approximation to proportional fairness. As a consequence of its proportional fairness, we show that this voting rule achieves $\Theta(\log m)$ distortion with respect to the Nash welfare, and selects a distribution that provides a $\Theta(\log m)$-approximation to the core, making it interesting for applications in participatory budgeting. For all three approximations, we show that $\Theta(\log m)$ is the best possible approximation.
\end{abstract}

\section{Introduction}
\label{sec:intro}

We consider the problem of designing voting rules that aggregate agents' ranked preferences and arrive at a collective decision with high social welfare and which is fair to all agents. 
Throughout, we focus on probabilistic voting rules, which take as input a preference profile of complete rankings of a set $A$ of $m$ alternatives and output a probability distribution over $A$. 

\crefname{theorem}{Thm.}{Thm.}
\newcommand{\tableref}[1]{\!{\small[#1]}}
\begin{table}[t]
\hspace{-10pt}
	\scalebox{0.885}{
		\renewcommand{\arraystretch}{1.1}
		\begin{tabular}{@{}llrlrlr@{}}
			\toprule
			& \multicolumn{2}{l}{Distortion, unit-sum} 
			& \multicolumn{2}{l}{Distortion, unit-range} 
			& \multicolumn{2}{l}{Proportional fairness} \\
			\midrule
			Stable Lottery Rule 
			& $\Theta(\sqrt{m})$ & \tableref{\cref{thm:distortion_ub}}
			& $\Theta(\sqrt{m})$ & \tableref{\cref{thm:distortion_ub}}
			& --- & \\
			Stable Committee Rule 
			& $\Theta(\sqrt{m})$  & \tableref{\cref{thm:distortion_ub_stable_committee}}
			& $\Theta(\sqrt{m})$ & \tableref{\cref{thm:distortion_ub_stable_committee}}
			& $O(\sqrt{m})$ & \tableref{\cref{thm:pf-scr}}
			\\
			Harmonic Rule (strategyproof) 
			& $\Theta(\sqrt{m \log m})$ &\tableref{\cref{thm:LB-harmonic-rule-utilitarian}}
			& $\smash{\Theta(m^{\frac23} \log^{\frac13} m)}$ & \tableref{\cref{thm:hr-approval}}
			& $\Theta(\sqrt{m \log m})$ &\tableref{\cref{thm:hr-pf}}
			\\
			\midrule
			Best possible 
			& $\Theta(\sqrt{m})$ & \hspace{-10pt}\tableref{BCHLPS 15}
			& $\Theta(\sqrt{m})$ & \tableref{\cref{thm:LB_approval}}
			& $\Theta(\log m) $ & \hspace{-15pt}\tableref{\cref{thm:UB-pf-logm}, \ref{thm:LB-pf-2}}
			\\
			Best possible, strategyproof
			& $\Theta(\sqrt{m \log m})$ &\tableref{BDG 18}
			& $\smash{\Theta(m^{\frac23})}$ & \tableref{FM 14}
			& $\Omega(\sqrt{m})$ & \tableref{\cref{thm:strategyproof-lb-pf}}
			\\
			\bottomrule
	\end{tabular}}
\caption{Overview of our results. Previously known results: $O(\sqrt{m \log m})$ upper bound on the distortion of the harmonic rule for unit-sum utilities~\citep{BCHL+15}, $\Omega(\sqrt{m})$~\citep{BCHL+15} and $\Omega(\sqrt{m \log m})$~\citep{bhaskar2018truthful} lower bounds on the distortion of non-strategyproof and strategyproof rules for unit-sum utilities, respectively, and $\Theta(m^{2/3})$ distortion of the best possible strategyproof rule for unit-range utilities~\citep{filos2014truthfulrange,lee2019maximization}.} 
	\label{tbl:results-table}
\end{table}

In order to evaluate the social welfare and fairness of voting rules, we build upon the framework of \textit{implicit utilitarian voting}~\citep{procaccia2006distortion}, which assumes that each agent $i$ has a cardinal utility function $u_i : A \to \bbR_{\ge 0}$ over alternatives, but reports only the induced ranking over alternatives to the voting rule. 
While in principle a voting rule could elicit the precise utility values, it is more common in the literature to ask for rankings. This makes for a simple elicitation protocol, which can ease the cognitive burden on agents (because they need not precisely determine their own utility values), and preserves the privacy of any agents who may not wish to reveal their exact utilities to a voting rule.

The implicit utilitarian framework allows us to quantify the efficiency of a given voting rule:
Given an input profile of rankings, we can measure  efficiency as the worst-case ratio between the social welfare of the optimal outcome and the social welfare of the outcome selected by the voting rule, where the worst case is taken over all utility functions consistent with the ordinal rankings reported to the voting rule. 
This quantity is known as \textit{distortion} and has been widely studied.
The existing literature commonly analyzes distortion for the class of \emph{unit-sum utilities}, in which each agent's total utility is normalized to $1$~\citep{BCHL+15,CNPS17,MPSW19,MPW20,filos2020distortion}. For deterministic voting rules, which place the entire probability mass of $1$ on a single alternative, it is known that the best possible distortion is $\Theta(m^2)$~\citep{CP11,CNPS17}. For probabilistic voting rules, which we are interested in, \citet{BCHL+15} prove a lower bound of $\Omega(\sqrt{m})$. We propose the first voting rule achieving the asymptotically optimal distortion of $O(\sqrt{m})$, matching the lower bound of \citet{BCHL+15} and resolving an important open question in this line of work. Our proof shows that the same rule is also optimal for \emph{unit-range} utilities (which are normalized to range between $0$ and $1$) with the same $O(\sqrt{m})$ distortion. This improves upon the previous best known distortion of $O(m^{2/3})$ \citep{lee2019maximization,filos2015social}. This $O(\sqrt{m})$ distortion of our rule is also optimal for the special case of \emph{approval utilities}, in which each agent has utility $1$ for a subset of alternatives and utility $0$ for the rest. This class corresponds naturally to approval voting but, to the best of our knowledge, has not been studied in the context of distortion.\footnote{Distortion for approval utilities makes sense in contexts where agents may find it easier to rank alternatives than to assign them approval utilities. For example, if the alternatives are budget divisions, a project leader would naturally rank the divisions by the amount of money allocated to their project. But the eventual utility depends on whether the money is enough to deliver the project or not, and the required amount may be unknown at the time of voting.}
Our rule can be computed in polynomial time.

Interestingly, while our new voting rule achieves low distortion (i.e., high social welfare), it internally aims for a fair outcome. Specifically, it uses tools from multi-winner voting for selecting a committee (a fixed-size subset of alternatives) that is \emph{representative}. Informally, as many agents as possible should have one of their highly ranked alternatives in the committee. There is an intuitive case for considering representative committees for achieving low distortion: Suppose a voting rule places little weight on the highly-ranked alternatives of a large group of agents. Then the voting rule may incur high distortion when those agents feel strongly about their preferences and all other agents are indifferent. This suggests that, at least in some settings, if one wants to be efficient, it pays to also be fair.

While we use fair committees as a means to achieve high social welfare, we are also interested in fairness as an end. We wish to achieve a notion of fairness defined for our single-winner setting. Specifically, we adapt a quantitative measure from network theory called \emph{proportional fairness} to the voting context. This measure is phrased in terms of agents' utility functions, and we combine it with the worst-case philosophy of distortion to obtain a way to measure the fairness of voting rules. Intuitively, for an outcome to do well with respect to proportional fairness, it cannot be the case that any large group of agents gets too little utility, where ``too little'' is a function of how large the group is and how easy it is to give high utility to the group.

If we knew the underlying agent utilities, we could compute a distribution that is $1$-proportionally fair. We show that given only a preference profile of rankings, there always exists a distribution that is $O(\log m)$-proportionally fair regardless of the underlying utility functions (consistent with the input rankings). Our existence proof uses the minimax theorem for zero-sum games. We show that our result is optimal because there are preference profiles for which every distribution has an approximation to proportional fairness that is no better than $\Omega(\log m)$ under some consistent utility functions. We then show that, given a preference profile, the projected subgradient descent algorithm can be used to compute a distribution with an (almost) optimal approximation to proportional fairness in polynomial time.

Proportional fairness is an interesting measure because voting rules that do well on it automatically do well on other fairness measures as well. For example, it is widely recognized that maximizing the \emph{Nash welfare} instead of the utilitarian welfare gives fairer outcomes (the Nash welfare of an outcome is the \emph{product} of agent utilities instead of the sum). We can define a version of distortion for the Nash welfare, and our rule for proportional fairness will guarantee $O(\log m)$ distortion for this objective, which we show to be best-possible.
Another fairness property is taken from the literature on \emph{participatory budgeting} (PB) \citep{FGM16a}. We can interpret a probabilistic voting rule as dividing a fixed budget between different projects, and agents vote by ranking those projects. Agents wish to see more money spent on the projects they rank higher. An important goal in PB is to provide proportional representation in that any $x\%$ of the population cannot find an allocation of $x\%$ of the budget which provides them a Pareto improvement (i.e., does not hurt any of them and strictly improves some). This aim can be formalized using the concept of the core. Our rule for proportional fairness selects an outcome that provides an $O(\log m)$-approximation to the core, which we show to be best-possible.

\Cref{tbl:results-table} provides an up-to-date account of the results on distortion for unit-sum and unit-range utilities as well as proportional fairness. It also includes known and new results for strategyproof rules which we discuss in \Cref{app:harmonic,sec:truthful-pf}.

\subsection{Related Work}

There are many papers that study the distortion of voting rules, beginning with the work of~\citet{procaccia2006distortion}, who analyze the distortion of many common voting rules.
\citet{CP11} also evaluate the distortion of prominent voting rules, but from the perspective of optimizing embeddings, which (perhaps randomly) map cardinal utilities to ordinal votes that voting rules take as input. Their work, together with that of \citet{CNPS17}, identifies the best possible distortion via deterministic voting rules to be $\Theta(m^2)$. \citet{BCHL+15} study probabilistic voting rules and derive a lower bound of $\Omega(\sqrt{m})$ on the optimal distortion for probabilistic voting rules with unit-sum utilities. They also design an artificial rule (tailored specifically to the unit-sum distortion context) which establishes an upper bound of $O(\sqrt{m} \log^*\! m)$.\footnote{$\log^*\! m$ is the \emph{iterated logarithm} of $m$, which is the number of times $\log$ needs to be applied to $m$ until the result is at most $1$.} Our $O(\sqrt{m})$ upper bound matches their lower bound and eliminates the $\log^*\! m$ gap.  

\citet{BCHL+15} also propose the \emph{harmonic rule} based on the harmonic scoring rule and show that it achieves $O(\sqrt{m \log m})$ distortion for unit-sum utilities. \citet{bhaskar2018truthful} point out that this voting rule is strategyproof (in expectation with respect to any consistent utility function), and prove that any strategyproof rule must incur at least $\Omega(\sqrt{m \log m})$ distortion, making the harmonic rule asymptotically optimal, subject to strategyproofness. Distortion subject to strategyproofness had first been studied by \citet{filos2014truthfulrange}, who consider unit-range utilities and prove that any strategyproof rule must incur at least $\Omega(m^{2/3})$ distortion. Their proof also implies this bound for approval utilities. \citet{lee2019maximization} proposed a strategyproof method that achieves a matching $O(m^{2/3})$ upper bound for unit-range utilities.\footnote{The method achieving this bound chooses, with probability $\frac12$ an alternative uniformly at random, and with probability $\frac12$ it chooses a voter uniformly at random and then chooses one alternative that this voter ranks in the top $m^{1/3}$ ranks uniformly at random. A detailed proof of the result by \citet{lee2019maximization} that this rule achieves $O(m^{2/3})$ distortion for unit-range utilities is presented by \citet[Section~2.3]{filos2015social}.}
In the appendix, we show that the harmonic rule achieves distortion $\Theta(m^{2/3} \log^{1/3}m)$ for approval and for unit-range utilities, matching the lower bound subject to strategyproofness up to a logarithmic factor. Using the techniques of \citet{filos2014truthfulrange} and \citet{bhaskar2018truthful}, we derive a lower bound of $\Omega(\sqrt{m})$ on our proportional fairness objective subject to strategyproofness, and show that the harmonic rule again matches this bound, up to a logarithmic factor. 

Implicit utilitarian voting can be seen as a protocol for reducing communication complexity by asking agents to report ordinal preferences in place of cardinal utilities, so it is natural to study the trade-off between the communication complexity (the number of bits of information each agent is asked to report) and the optimal distortion achievable. \citet{MPSW19,MPW20} characterize the Pareto frontier of this tradeoff, showing that in order to achieve distortion $d$, probabilistic voting rules require agents to communicate only $\tilde{\Theta}(m/d^3)$ bits of information whereas deterministic voting rules require $\tilde{\Theta}(m/d)$ bits, establishing probabilistic rules as superior in this context.
\citet{amanatidis2021peeking} considered making a few value queries (asking agents to report their utility for an alternative) or comparison queries (asking agents to report whether the ratio of their utilities for two alternatives is at least a threshold) on top of their reported ordinal preferences. They prove that asking only $O(\log^2 m)$ value queries or $O(\log^2 m)$ comparison queries is sufficient to achieve constant distortion. 

Going beyond single-winner voting, \citet{CNPS17} study distortion (and another closely related objective called regret) for multi-winner voting, where the goal is to select a committee of $k$ alternatives for a given size $k$. They assume that the utility of an agent for a committee is the maximum utility of the agent for any alternative in the committee. They prove that the optimal distortion of deterministic rules is $\Theta(1+m (m-k)/k)$, implying an optimal distortion of $\Theta(m^2)$ for deterministic single-winner voting. For probabilistic rules, they leave a gap of $\Theta(m^{1/6})$ between their upper and lower bounds for the optimal distortion. Recently, \citet{borodin2022distortion} close this gap by building upon our work. They extend our single-winner rule with $O(\sqrt{m})$ distortion to multi-winner voting and prove that it achieves the optimal distortion of $\Theta(\min(\sqrt{m},m/k))$. 

\citet{BNPS21} study participatory budgeting, which is an extension of multi-winner voting in which each alternative has a cost and the goal is to find a subset of alternatives with total cost at most a given budget. They focus on a different utility model, in which the utility of an agent for a set of alternatives is the sum of her utilities for the alternatives in the set. They compare four protocols for eliciting agent preferences and prove that while ranked preferences lead to $O(\sqrt{m} \cdot \log m)$ distortion with probabilistic aggregation, threshold approval votes, which ask agents to identify alternatives for which their utility is at least a specified threshold, lead to a significantly lower distortion of $O(\log^2 m)$. \citet{bhaskar2018truthful} show that the near-optimal $O(\sqrt{m} \cdot \log m)$ distortion for participatory budgeting with ranked preferences can in fact be obtained via a strategyproof voting rule, establishing that strategyproofness comes at minimal cost even in this general model.

In all these papers, agents are modeled to have \emph{normalized utilities} for alternatives. Initiated by the work of \citet{anshelevich2018approximating}, a large recent literature about \emph{metric distortion} instead models agents as having \emph{costs} for alternatives. This literature makes the assumption that the cost of an agent for an alternative is the distance between them in an underlying metric space, and aims to approximate the utilitarian social cost (i.e., the sum of agent costs)~\citep{anshelevich2016blind,anshelevich2017randomized,anshelevich2018approximating,munagala2019improved,kempe2020communication,caragiannis2022metric}. It turns out that the metric structure allows significantly better distortion bounds: the best distortion of deterministic rules is $3$~\citep{gkatzelis2020resolving,kizilkaya2022pluralityveto} (compared to $\Theta(m^2)$ in the non-metric setting) and that for probabilistic rules is between $2.1126$ and $2.753$~\citep{charikar2022metric,charikar2023breaking} (compared to $\Theta(\sqrt{m})$ distortion in the non-metric setting). Note that probabilistic rules are superior to deterministic rules in the metric setting as well.

Fairness of single-winner voting rules has received less attention than distortion. For probabilistic voting rules, fairness has been studied in a series of papers that interpret the output distribution as a division of a budget. Most work has studied this in a model with known approval utilities of the agents~\citep{BMS05a,ABM17a,Dudd15a,BBPS21a}. \citet{ordinalportioning} study probabilistic voting rules which take as input ranked preferences, and then convert those preferences into utilities using a fixed scoring vector (such as Borda). The rules then maximize the Nash welfare (the geometric mean of agent utilities) or the egalitarian welfare (the minimum agent utility) and its leximin refinement. Note that in our work the utilities are unknown. They prove that Nash-welfare-based rules satisfy the \emph{SD-core}. This is a weaker axiom than the core that we introduce in \Cref{sec:prelim}, which in the terminology of \citet{ABBB15a} could be called the \emph{strong SD-core}. We note that SD-core implies no better than an $m$-approximation of our (strong) core (for example, random dictatorship satisfies SD-core and is in the $m$-approximate core), whereas we achieve an $O(\log m)$-approximate core. In a model where voters report their utilities, \citet{FGM16a} investigate the core and propose a polynomial-time algorithm for finding an outcome in the core via the so-called Lindahl equilibrium. Note that they do not require an approximation to the core because utilities are known. They also point out connections to proportional fairness. 

Fairness in voting has been studied in detail for deterministic multi-winner voting rules. Various fairness notions have been studied that require every group of agents to have representation in the committee, with larger and more cohesive groups having better representation. This includes notions such as justified representation (JR), extended justified representation (EJR)~\citep{aziz2017justified}, proportional justified representation (PJR)~\citep{sanchez2017proportional}, full justified representation (FJR)~\citep{PPS21} and the proportionality degree~\citep{skowron2021proportionality}. \citet{cheng2020group} prove that there always exists a distribution over committees that satisfies a stronger fairness notion called \emph{stability}; this is the main tool we use to achieve $O(\sqrt{m})$ distortion for single-winner voting. \citet{jiang2020approximately} derandomize this result to prove that there always exists a committee satisfying $32$-approximate stability; we show that this derandomized result can be used to achieve $O(\sqrt{m})$ distortion with respect to the Nash welfare, but we are able to improve on that bound to achieve $\Theta(\log m)$ distortion using the minimax theorem (which is best-possible). \citet{FMS18a} study a more general model of public goods and achieve different approximations to the core under various constraints on feasible outcomes. 

\section{Preliminaries}
\label{sec:prelim}
For $t \in \bbN$, we write $[t] = \set{1,\ldots,t}$. For a set $A$, let $\Delta(A)$ be the set of probability distributions $\xx$ over $X$. For $a \in A$, we write $\xx(a)$ for the probability that $\xx$ places on $a$, and for a set $A' \subseteq A$, we write $\xx(A') = \sum_{a'\in A'} \xx(a')$.

We repeatedly use the inequality of arithmetic, geometric, and harmonic means (AM-GM-HM inequality) which states that for all $a_1, \dots, a_t > 0$, we have $\frac{1}{t} \sum_{i = 1}^t a_i \ge \sqrt[t]{a_1  a_2 \cdots a_t} \ge \frac{t}{\frac{1}{a_1} + \cdots \frac{1}{a_t}}$.

\paragraph{Voting.} 
Let $N$ be a set of $n$ agents and $A$ be a set of $m$ alternatives. For $k \in [m]$, let $\power_k(A)$ denote the set of all subsets of $A$ of size $k$. Each agent $i \in N$ submits a \emph{preference ranking} over the alternatives, encoded by a bijective rank function $\pref_i : A \to [m]$. For example, if $\pref_i(a) = 1$, then $a$ is the most-preferred alternative for agent $i$. We use $a \succ_i a'$ to denote $\pref_i(a) < \pref_i(a')$ (agent $i$ ranks $a$ strictly above $a'$) and $a \succeq_i a'$ to denote $\pref_i(a) \le \pref_i(a')$. We refer to the collection $\profile = (\pref_i)_{i \in N}$ as a \emph{preference profile}. A \emph{probabilistic voting rule} $f$ (which we usually just call a \emph{voting rule}) is a function that takes a preference profile $\profile$ as input and outputs a distribution over alternatives. The output of a voting rule can be interpreted as a randomized selection of alternatives, but also as a division of some divisible resource (such as time or a budget) between the alternatives.

\paragraph{Utilities.} A \emph{utility function} $u : A \to \bbR_{\ge 0}$ assigns a non-negative utility to each alternative. We can extend $u$ to also assign utility values to distributions $\xx \in \Delta(A)$ over alternatives by setting $u(\xx) = \E_{a \sim \xx} u(a)$. We assume that when agents submit ranked preferences, they have more expressive underlying cardinal preferences. Given a preference profile $\profile$, we say that a utility function $u_i$ for agent $i$ is \emph{consistent} with her preference ranking if for all $a,a' \in A$ such that $a \succ_i a'$, we have  $u_i(a) \ge u_i(a')$. Note that we allow alternatives to have equal utility, and then the agent can break ties arbitrarily when submitting a preference ranking. We refer to a collection $\utilities = (u_i)_{i \in N}$ as a \emph{utility profile}. We use the notation $\utilities \induces \profile$ to indicate that $u_i$ is consistent with $\pref_i$ for each agent $i$. Note that voting rules have access to the preference profile but not to the utility profile.

\paragraph{Utility classes.} Let $\utilityclass^{\all}$ denote the class of all possible utility functions. We also study the following standard restricted utility classes.
\begin{itemize}
	\item $\utilityclass^{\unitsum}$ is the class of \emph{unit-sum} utility functions $u$ satisfying $\sum_{a \in A} u(a) = 1$. 
	\item $\utilityclass^{\unitrange}$ is the class of \emph{unit-range} utility functions $u$ satisfying $\max_{a \in A} u(a) = 1$.\footnote{Some definitions of unit-range utilities require $\min_{a \in A} u(a) = 0$ in addition, but this is not necessary for our results.} 
	\item $\utilityclass^{\approval}$ is the class of \emph{approval} utility functions $u$ satisfying $u(a) \in \set{0,1}$ for all $a \in A$ and $u(a) = 1$ for at least one $a \in A$.
\end{itemize} 

We introduce a new class of \emph{balanced} utility functions, where the highest utility intensity that can be expressed is at most $1$, and where the total utility of the utility function is at least $1$.
\begin{itemize} 
	\item $\utilityclass^{\balanced}$ is the class of utility functions $u$ satisfying $u(a) \le 1$ for all $a \in A$ and $\sum_{a \in A} u(a) \ge 1$. 
\end{itemize} 

Note that $\utilityclass^{\unitsum} \subseteq \utilityclass^{\balanced}$ and $\utilityclass^{\approval} \subseteq \utilityclass^{\unitrange} \subseteq \utilityclass^{\balanced}$. 
Our main upper bound for distortion (\Cref{thm:distortion_ub}) will hold for the entire class of balanced utility functions. \\

In this work, we focus on two metrics for evaluating voting rules: distortion, which is a measure of social welfare, and proportional fairness, which is a measure of fairness.

\subsection{Distortion}\label{sec:prelim-dist}
Given the utility profile $\utilities$, the \emph{utilitarian welfare} of a distribution over alternatives $\xx \in \Delta(A)$ is defined as $\SW(\xx,\utilities) = \sum_{i \in N} u_i(\xx)$.

If one could observe the underlying utilities, an argument dating back to \citet{bentham} suggests that picking the alternative maximizing the utilitarian welfare is the best choice. However, a voting rule is allowed to observe only the preference profile $\profile$, thus obtaining partial information about the utility profile $\utilities$. In this case, we measure the efficiency of the voting rule by the worst-case approximation ratio it achieves for maximizing the utilitarian welfare. 

\begin{definition}[Distortion]\label{def:distortion}
Given a utility profile $\utilities$, the \emph{distortion} of a distribution $\xx \in \Delta(A)$ is the ratio between the highest possible social welfare and the social welfare of $\xx$ under $\utilities$:
\[
\D(\xx, \utilities) = \frac{\max_{\yy \in \Delta(A)} \SW(\yy,\utilities)}{\SW(\xx,\utilities)}.
\]
The distortion of $\xx$ on a preference profile $\profile$ for a utility class $\utilityclass$ is obtained by taking the worst case over all utility profiles $\utilities \in \utilityclass^n$ consistent with $\profile$. 
\[
\D(\xx, \profile, \utilityclass) = \textstyle\sup_{\utilities \in \utilityclass^n :\,\utilities \induces \profile}\, \D(\xx,\utilities).
\]
Given a number $m$ of alternatives, the distortion of a voting rule $f$ for utility class $\utilityclass$ is $\D_{m}(f, \utilityclass) = \sup_{\profile} \D(f(\profile),\profile,\utilityclass)$, where the supremum is taken over all preference profiles $\profile$ with $m$ alternatives and any number of agents.
\end{definition}

\begin{table}[ht]
	\centering
	\begin{tabular}{ll c@{$\,$}c@{$\,$}c@{$\,$}c@{$\,$}c c@{$\,$}c@{$\,$}c@{$\,$}c@{$\,$}c c@{$\,$}c@{$\,$}c@{$\,$}c@{$\,$}c}
		\toprule
		&&  \multicolumn{5}{c}{Agent $i_1$} &  \multicolumn{5}{c}{Agent  $i_2$} & \multicolumn{5}{c}{Agent  $i_3$} \\
		\midrule
		Preferences & $\profile$ &
		$a_1$ & $\succ$ & $a_2$ & $\succ$ & $a_3$ &
		$a_2$ & $\succ$ & $a_1$ & $\succ$ & $a_3$ &
		$a_1$ & $\succ$ & $a_3$ & $\succ$ & $a_2$
		\\
		\multirow{2}{*}{Utilities} & $\utilities_{1}$ &
		$\nicefrac{1}{2}$ & $,$ & $\nicefrac{1}{3}$& $,$ & $\nicefrac{1}{6}$ & $ \nicefrac{1}{2}$& $,$ & $ \nicefrac{1}{2}$& $,$ & $0$ & $ \nicefrac{1}{3}$& $,$ & $ \nicefrac{1}{3}$& $,$ & $ \nicefrac{1}{3}$\\
		& $\utilities_{2}$ &
		$\nicefrac{1}{2}$ & $,$ & $\nicefrac{1}{2}$& $,$ & $0$ & $1$& $,$ & $0$& $,$ & $0$ & $\nicefrac{1}{3}$& $,$ & $\nicefrac{1}{3}$& $,$ & $\nicefrac{1}{3}$\\
		\bottomrule
	\end{tabular}
	\caption{A preference profile  with three agents and three alternatives along with two consistent unit-sum utility profiles. Utilities are sorted according to agents' preference rankings.}
	\label{tbl:example-distortion}
\end{table}

\begin{example}
	\label{exm:distortion}
	\Cref{tbl:example-distortion} shows a preference profile with three agents and three alternatives. Consider the distribution $\xx = (a_1: \nicefrac{1}{2},\, a_2: \nicefrac{1}{4},\, a_3: \nicefrac{1}{4})$. Let us evaluate its distortion under the two utility profiles given in the table. 
	\begin{itemize}
		\item For utility profile $\utilities_{1}$, the social welfare of $\xx$ is $\SW(\xx, \utilities_{1}) = \nicefrac{3}{8} + \nicefrac{3}{8} + \nicefrac{1}{3} = \nicefrac{13}{12}$. 
		For $\utilities_1$, the optimal outcome is $\yy = (a_1 : 1)$ with $\SW(\yy,\utilities_1) = \nicefrac{4}{3}$. 
		Hence, $\D(\xx, \utilities_{1}) = \frac{\nicefrac{4}{3}}{\nicefrac{13}{12}} \approx 1.23$.
		\item For utility profile $\utilities_{2}$, the social welfare of $\xx$ is $\SW(\xx, \utilities_{2}) = \nicefrac{3}{8} + \nicefrac{1}{4} + \nicefrac{1}{3} = \nicefrac{23}{24}$. 
		For $\utilities_2$, the optimal outcome is $\yy = (a_2 : 1)$ with $\SW(\yy,\utilities_2) = \nicefrac{11}{6}$. 
		Hence, $\D(\xx, \utilities_{2}) = \frac{\nicefrac{11}{6}}{\nicefrac{23}{24}} = \nicefrac{44}{23} \approx 1.91$.
	\end{itemize}
	Thus, under $\utilities_1$, it is possible to obtain 23\% more social welfare than $\xx$, and under $\utilities_2$, it is possible to obtain 91\% more. Using a simple linear program, one can check that $\utilities_2$ is the worst case for utility profiles from $\utilityclass^{\unitsum}$, so $\D(\xx, \profile, \utilityclass^{\unitsum}) = \nicefrac{44}{23} \approx 1.91$. Using a more sophisticated linear program \citep{BCHL+15}, one can find the distribution with the lowest possible distortion for unit-sum utilities, which is $\xx^* \approx (a_1: 0.5882, a_2: 0.4118, a_3: 0)$ with distortion of 
	about 1.54.
\end{example}

As we mention in \Cref{exm:distortion}, given a preference profile $\profile$, one can find a distribution $\xx$ minimizing $\D(\xx, \profile, \utilityclass^{\unitsum})$ by solving a linear program proposed by \citet{BCHL+15}. Their approach works for any utility class that is described by linear constraints, so it can be used to find instance-optimal distributions for unit-sum, for unit-range, and for balanced utilities. However, approval utilities are not described by linear constraints (since we need to enforce integrality). Still, we show in \Cref{lem:approval-unit-range} in the appendix that the distribution minimizing distortion for unit-range utilities also minimizes distortion for approval utilities, so instance-optimal distributions for approval utilities can still be found using a linear program.

We have defined the distortion for a class of utility functions $\utilityclass$ by taking the worst case over all utility profiles $\utilities$ in which the utility function $u_i$ of every agent $i$ belongs to $\utilityclass$. Most naturally, one would like to analyze the distortion for the class of all utility functions $\utilityclass^{\all}$. However, the worst-case distortion for this class is degenerate: the rule that always selects the uniform distribution has $O(m)$ distortion, and it is easy to see that any rule has at least $\Omega(m)$ distortion (by considering utility profiles where some agents care a lot and others not at all). Thus, without some additional restrictions on cardinal utilities (such as unit-sum or unit-range), it turns out that ordinal preferences do not provide significant information about utilitarian welfare. 

\subsection{Nash Welfare Distortion} 

Distortion is typically defined with respect to utilitarian welfare, but the same principle can be applied to other welfare functions. We will in particular study \emph{Nash welfare} ($\NW$), which is the geometric mean of agent utilities: $\NW(\xx,\utilities) = \left( \prod_{i \in N} u_i(\xx) \right)^{\nicefrac{1}{n}}$. We can define the distortion $\D_{m}^{\NW}(f,\utilityclass)$ of a voting rule $f$ for Nash welfare by replacing the utilitarian welfare $\SW$ in \Cref{def:distortion} by $\NW$. 

Nash welfare is sometimes viewed as a combined measure of efficiency and fairness. It measures efficiency in a Pareto sense (if everyone's utility increases then so does Nash welfare), and it measures fairness because if some agent has very low utility then this has a strong negative impact on overall Nash welfare.
Remarkably, the Nash welfare is \emph{scale invariant}, i.e., multiplying the utility function of an agent by some factor does not change the comparison between the Nash welfare of two distributions over alternatives. Hence, we have that $\D_{m}^{\NW}(f,\utilityclass^{\all}) = \D_{m}^{\NW}(f,\utilityclass^{\unitsum}) = \D_{m}^{\NW}(f,\utilityclass^{\unitrange})$ for every voting rule $f$.

\subsection{Core}

When we view a distribution $\xx$ as a division of a budget between the alternatives, the core is a fairness axiom that intuitively guarantees every group of agents an influence proportional to its size, provided the agents in the group have similar enough preferences.

Let $\alpha \ge 1$. We will define an $\alpha$-approximate notion of the core which coincides with the standard version when $\alpha = 1$. Similar $\alpha$-approximations to the core have been studied in discrete budgeting settings \citep{FMS18a,peters2020laminar,munagala2022approximate}. A distribution over alternatives $\xx \in \Delta(A)$ is said to be in the \emph{$\alpha$-core} with respect to utility profile $\utilities$ if there is no subset of agents $S$ and distribution over alternatives $\yy \in \Delta(A)$ such that 
\[
\tfrac{|S|}{|N|} \cdot u_i(\yy) \ge \alpha \cdot u_i(\xx) \,\,
\text{for every agent $i \in S$,}
\]
and at least one of these inequalities is strict.\footnote{Equivalently, this condition requires that there is no set $S \subseteq N$ and \emph{partial} distribution $\yy: A \to [0,1]$ with $\sum_a \yy(a) = |S|/n$ such that $u_i(\yy) \ge \alpha \cdot u_i(\xx)$} for every agent $i \in S$ and at least one of these inequalities is strict \citep{FGM16a,FMS18a}.
For any utility profile $\utilities$, the distribution maximizing Nash welfare is in the 1-core \citep{FGM16a}.
Given a preference profile $\profile$, we say that a distribution $\xx$ is in the \emph{universal $\alpha$-core} if $\xx$ is in the the $\alpha$-core with respect to every utility profile $\utilities$ consistent with $\profile$.
A voting rule $f$ is said to be in the universal $\alpha$-core if, for every preference profile $\profile$, $f(\profile)$ is in the universal $\alpha$-core.

The core notion is inspired by cooperative game theory, where it is seen as a stability notion: if a group of agents is not treated fairly, then those agents can leave the system in order to use their fraction of the budget in a preferable way. The core is typically defined in settings where agents' utility functions are known; our notion is phrased for the case where the rule has access to only ordinal information. Note that to be in the universal $\alpha$-core, a rule needs to avoid deviations for all consistent utilities. This makes sense as a conservative stability notion because agents will presumably make their decision to leave based on their actual utilities.

While $1$-core can be achieved when exact utilities are known (see \Cref{sec:prelim-pf}), it is easy to see that no rule can satisfy $1$-core given only the ordinal preferences. For example, consider the preference profile from \Cref{exm:distortion} with preferences ($a_1 \succ a_2 \succ a_3$, $a_2 \succ a_1 \succ a_3$, $a_1 \succ a_3 \succ a_2$). For the utility profile where each agent approves just their top alternative, there is a unique distribution $\xx$ that satisfies the 1-core, namely $\xx = (a_1 : \nicefrac{2}{3}, a_2 : \nicefrac{1}{3}, a_3 : 0)$. This is because if $\xx(a_1)$ was any lower, then the first and third agents could deviate; if $\xx(a_2)$ was lower, then the second agent could deviate. However, $\xx$ fails the 1-core if we change the utility profile so that the second agent gives the same utility to all alternatives. For that utility profile, all three agents can deviate together by proposing to place the entire budget on $a_1$.

\subsection{Proportional Fairness}\label{sec:prelim-pf}
We have now seen two notions that are connected to fairness (Nash welfare and the core). 
A third such notion is \emph{proportional fairness}, which was first proposed in communication networks~\citep{kelly1998rate} but is easily adapted to social choice more generally. 
This is a quantitative way of measuring the fairness of a distribution.
As we will see, proportional fairness is intimately connected to the other two notions.

\begin{definition}[Proportional Fairness]
\label{def:pf}
Let $\xx \in \Delta(A)$ be a distribution over alternatives.
Given a utility profile $\utilities$, we write
\begin{equation}\label{eqn:pf}
\PF(\xx,\utilities) = \max_{\yy \in \Delta(A)} \frac{1}{n} \sum_{i \in N} \frac{u_i(\yy)}{u_i(\xx)} = \max_{a \in A}\, \frac{1}{n} \sum_{i \in N} \frac{u_i(a)}{u_i(\xx)}.\footnote{This is the maximum possible average multiplicative increase in agent utilities when moving from $\xx$ to any other $\yy$. The second transition in \eqref{eqn:pf} holds because $\frac{1}{n} \sum_{i \in N} u_i(\yy)/u_i(\xx)$ is linear in $\yy$.}
\end{equation}
\end{definition}

For every utility profile $\utilities$, there exists a distribution $\xx$ with $\PF(\xx, \utilities) = 1$; in fact, the distribution that maximizes Nash welfare with respect to $\utilities$ has this property \citep[e.g.,][Sec.~2.2]{FMS18a}. This is the lowest possible value because if we take $\yy = \xx$ in the definition of $\PF(\xx, \utilities)$, we obtain a value of 1. To illustrate why proportional fairness is a measure of fairness, we can note that if $\xx$ is a distribution such that $u_i(\xx) = 0$ for some agent $i \in N$, then $\PF(\xx, \utilities) = \infty$, which we can see by taking any $\yy$ for which $u_i(\yy) > 0$.
Thus, an $\alpha$-proportionally fair distribution, with $\alpha$ not too high, guarantees to every agent a base level of utility compared to what the agent can receive in any other distribution. In particular, no agent's preferences can be completely ignored.

\begin{definition}[Distortion of Proportional Fairness]
Given a preference profile $\profile$ and a utility class $\utilityclass$, the distortion with respect to proportional fairness of a distribution $\xx$ is
\[
\D^{\PF}(\xx,\profile,\utilityclass) = \textstyle\sup_{\utilities \in \utilityclass^n :\,\utilities \induces \profile}\, \displaystyle \frac{\PF(\xx,\utilities)}{\min_{\yy \in \Delta(A)} \PF(\yy, \utilities)} = \textstyle\sup_{\utilities \in \utilityclass^n :\,\utilities \induces \profile}\, \PF(\xx,\utilities),
\] 
where the last transition holds because the denominator in the middle expression is always 1, as we just discussed.

Given a number $m$ of alternatives, the distortion with respect to proportional fairness of a voting rule $f$ (for  utility class $\utilityclass$) is obtained by taking the worst case over all preference profiles $\profile$ with $m$ alternatives and any number of agents, that is, $\D^{\PF}_{m}(f,\utilityclass) = \sup_{\profile} \PF(f(\profile),\profile,\utilityclass)$.
\end{definition}

Like the Nash welfare, proportional fairness is also scale invariant.
Hence, we have $\D^{\PF}_m(f,\utilityclass^{\all}) = \D^{\PF}_m(f,\utilityclass^{\unitsum}) = \D^{\PF}_m(f,\utilityclass^{\unitrange})$ for every voting rule $f$. Our results for proportional fairness all hold with respect to $\utilityclass^{\all}$, so we drop it from the notation and simply use $\D^{\PF}(\xx,\profile)$ and $\D^{\PF}_m(f)$.

\begin{example}
Consider the profile in \Cref{tbl:example-distortion}, previously discussed in \Cref{exm:distortion}. For the distribution $\xx = (a_1: \nicefrac{1}{2},\, a_2: \nicefrac{1}{4},\, a_3: \nicefrac{1}{4})$ and utilities $\utilities_{1}$, we have
$\PF(\xx, \utilities_{1}) = \frac13 \max \{\nicefrac{11}{3}, \nicefrac{29}{9},  \nicefrac{13}{9}\} = \nicefrac{11}{9} \approx 1.22$.
For utilities $\utilities_{2}$, we have $\PF(\xx, \utilities_{2}) = \frac13\max \{\nicefrac{7}{3}, \nicefrac{19}{3},  1\} = \nicefrac{19}{9} \approx 2.11$.
Using \Cref{lem:pf-approval-simplification} in \Cref{sec:fairness}, we can check that $\utilities_{2}$ is the worst-case utility profile for distribution $\xx$, so that $\PF(\xx, \utilities) \le   \nicefrac{19}{9} \approx 2.11$ for all utility profiles $\utilities \in \utilityclass^{\all}$. Hence $\D^{\PF}(\xx, \profile) = \nicefrac{19}{9}$.

Using the techniques described in \Cref{sec:fairness-computation}, we can establish that the optimal distribution with respect to proportional fairness is $\xx^* \approx (a_1: 0.586,\, a_2: 0.414,\, a_3: 0)$, and $\PF(\xx^*, \utilities) \le 1.472$ for all utility profiles $\utilities \in \utilityclass^{\all}$.\footnote{On this small example, one can find this optimum distribution $\xx^*$ by hand after noting that $a_3$ must receive probability 0. One derives $\xx^* = \{a_1 : 2 - \sqrt{2}, a_2 :\sqrt{2} - 1, a_3 : 0\}$ with $\D^{\PF}(\xx^*,\profile,\utilityclass^{\all}) = 1 + \sqrt{2}/3$.}
\end{example}

An appealing strength of proportional fairness is that it is related to other fairness properties of interest. In particular, an $\alpha$-proportionally fair voting rule is also in the universal $\alpha$-core, and has a distortion with respect to Nash welfare of at most $\alpha$.

\paragraph{Proportional fairness $\Rightarrow$ the core.} 
The following is a well-known relation between proportional fairness and the core.\footnote{This result has not been explicitly stated, but essentially the same proof is frequently used to show that distributions maximizing Nash welfare lie in the core [e.g., \citealp{FMS18a}, Section 2.2; \citealp{ABM17a}, Theorem 3].}

\begin{proposition}\label{prop:pf-core}
For each $m$ and every $\alpha \ge 1$, if $\D_m^{\PF}(f) \le \alpha$, then $f$ is in the universal $\alpha$-core. 
\end{proposition}
\begin{proof}
We prove that if $f$ violates the universal $\alpha$-core, then its distortion with respect to proportional fairness is at least $\alpha$. Suppose there exists a consistent pair of utility profile $\utilities$ and preference profile $\profile$ such that $\xx = f(\profile)$ is not in the $\alpha$-core with respect to $\utilities$. Then there is a subset of agents $S$ and a distribution over alternatives $\yy \in \Delta(A)$ such that $\frac{|S|}{n} \cdot u_i(\yy) \ge \alpha \cdot u_i(\xx)$ (i.e., $\frac{u_i(\yy)}{u_i(\xx)} \ge \alpha \cdot \frac{n}{|S|}$) for every agent $i \in S$ and at least one of these inequalities is strict. Hence,  
\[
\sum_{i \in S} \frac{u_i(\yy)}{u_i(\xx)} > \alpha \cdot n 
\quad\Rightarrow\quad
\frac{1}{n} \sum_{i \in N} \frac{u_i(\yy)}{u_i(\xx)} \ge \frac{1}{n} \sum_{i \in S} \frac{u_i(\yy)}{u_i(\xx)} > \alpha,
\]
showing that $\D_m^{\PF}(f) > \alpha$.
\end{proof}

\paragraph{Proportional fairness $\Rightarrow$ distortion with respect to the Nash welfare.} 
It is also well-known that proportional fairness is an upper bound on the approximation of (i.e., distortion with respect to) the Nash welfare.\footnote{Observations to this effect can be found, for example, in Appendix D of \citet{CKM+19a} and in the derivation of Equation (2) in \citet{inoue2022additive}.}

\begin{proposition}\label{prop:pf-nash}
For every voting rule $f$, we have $\D_m^{\NW}(f,\utilityclass^{\all}) \le \D_m^{\PF}(f,\utilityclass^{\all})$. 
\end{proposition}
\begin{proof}
This holds because for any pair of distributions over alternatives $\xx,\yy \in \Delta(A)$ and utility profile $\utilities$, we have 
\[
\frac{\NW(\yy,\utilities)}{\NW(\xx,\utilities)} = \left( \prod_{i \in N} \frac{u_i(\yy)}{u_i(\xx)} \right)^{1/n} 
\!\!\!\le 
\frac{1}{n} \sum_{i \in N} \frac{u_i(\yy)}{u_i(\xx)},
\]
by the inequality of arithmetic and geometric means.
\end{proof}

\subsection{The Minimax Theorem}

In several places, we will use some basic elements of the theory of zero-sum games and the minimax theorem. Recall that if $X \subseteq \bbR^n$ is a convex set and $f: X \to \bbR$ is a function, then $f$ is \emph{convex} if for all $\xx_1,\xx_2 \in X$ and all $0 \le \lambda \le 1$, we have $f(\lambda \xx_1 + (1 - \lambda) \xx_2) \le \lambda f(\xx_1) + (1 - \lambda) f(\xx_2)$. Further, $f$ is \emph{concave} if $-f$ is convex. For example, $f(x) = x^2$ is convex and $f(x) = \log x$ is concave; linear functions are both convex and concave.

\begin{theorem}[Minimax Theorem, \citealp{vNeu28a}]
\label{thm:minimax}
	Let $X \subseteq \bbR^n$ and $Y \subseteq \bbR^m$ be compact convex sets. Let $f : X \times Y \to \bbR$ be a continuous function that is concave in its first argument and convex in its second argument (that is, $f(\cdot, \yy)$ is concave for each fixed $\yy \in Y$ and $f(\xx, \cdot)$ is convex for each fixed $\xx \in X$). Then
	\[
	\max_{\xx\in X} \min_{\yy\in Y} f(\xx,\yy)
	=
	\min_{\yy\in Y} \max_{\xx\in X} f(\xx,\yy).
	\]
\end{theorem}

We can interpret this theorem as a statement about a two-player zero-sum game between a \emph{player} and an \emph{adversary}. The player can choose a strategy $\xx$ from the set $X$ while aiming to maximize the value $f(\xx,\yy)$, and the adversary can choose $\yy\in Y$ aiming to minimize the value. The minimax theorem states that (under certain convexity conditions) it does not matter in which order the players make their moves. In our applications, we have $X = \Delta(S_1)$ and $Y = \Delta(S_2)$ for some finite sets $S_1$ and $S_2$ of \emph{pure strategies}, so that $X$ and $Y$ are sets of \emph{mixed strategies}. In this case, the function~$f$ typically encodes an expected payoff, $f(\xx,\yy) = \E_{s_1\sim\xx, s_2\sim\yy}[g(s_1,s_2)]$ for some $g : S_1 \times S_2 \to \bbR$. Such an $f$ is linear in both arguments and hence satisfies the conditions of the minimax theorem. In our results about proportional fairness, we will need the full strength of the minimax theorem, allowing for functions $f$ that are not linear in both arguments. The equal value of the max-min and min-max expressions is known as the \emph{value} of the zero-sum game.

\newpage
\section{Distortion}
\label{sec:distortion}

We begin by aiming to achieve low distortion with respect to utilitarian social welfare. \citet{BCHL+15} consider unit-sum utilities and show that any rule must incur distortion at least $\Omega(\sqrt m)$. They also construct an intricate and artificial voting rule that achieves distortion $O(\sqrt{m} \log^*\!m)$, thus leaving a tiny gap. They also present a more natural voting rule that achieves distortion $O(\sqrt{m \log m})$, which we call the \emph{harmonic rule} $\HR$. It is based on the harmonic scoring rule, according to which each agent $i$ gives $1/k$ points to the alternative she ranks in the $k$-th position. Given a preference profile $\profile$, the \emph{harmonic score} of an alternative $a$ is $\harm(a) = \sum_{i\in N} 1/\pref_i(a)$. Now, with probability $\frac12$, the harmonic rule chooses an alternative uniformly at random, and with probability $\frac12$, it chooses an alternative $a$ with probability proportional to $\harm(a)$. Note that the harmonic scores of all alternatives sum to $nH_m$ where $H_m = 1 + \frac12 + \cdots + \frac1m$. Thus if $\xx = \HR(\profile)$ then
\[
	\xx(a) = \frac{1}{2m} + \frac{\harm(a)}{2nH_m} \,\, \text{for all $a \in A$.}
\]

In this section, we introduce a new rule that achieves distortion $O(\sqrt m)$, which is optimal up to a constant factor. This rule is based on concepts from cooperative game theory and from the theory of committee selection, and can be computed in polynomial time. Our rule turns out to have robustly good performance, in that its distortion remains low for other utility classes and other welfare functions. We will compare it throughout to the harmonic rule.

\subsection{Stable Lotteries}\label{sec:stable}

Below the hood, our new voting rule is based on \emph{multi-winner voting}, also known as \emph{committee selection}, which concerns the well-studied problem of selecting a committee $X \subseteq A$ of $k$ alternatives, based on the agents' preferences over the alternatives \citep{FSST17a}. One goal of the literature on multi-winner voting is to identify \emph{representative} committees, where as many agents as possible are represented in the committee, in the sense that one of their highly-ranked alternatives is included \citep{ChCo83a}. This is a type of fairness consideration and related to the idea of proportional representation which is particularly well-developed in the context of approval utilities \citep{lackner2020approval}.

Representative committees are interesting in the distortion context due to the following intuition: if a voting rule places very little weight on alternatives that are highly ranked by many agents, then the rule is in danger of incurring high distortion, because those unrepresented agents may feel strongly about their high-ranked alternatives, while others may be more or less indifferent.

For ranked preferences, a recently studied representation axiom is \emph{(local) stability} 
\citep{aziz2017condorcet, cheng2020group}. This axiom is based on the idea that a group of $\frac{n}{k}$ agents should be able to decide over one of the $k$ slots in the committee. Formally, for a committee $X$ with $|X| = k$ and an alternative $a^*$, write $V(a^*,X) = |\{ i \in N : a^* \succ_i X\}|$ for the number of agents who prefer $a^*$ to all alternatives in the committee. We say that $X$ is \emph{stable} if for all alternatives $a^* \not\in X$, we have $V(a^*,X) < \frac{n}{k}$. Such a committee is stable in a sense familiar from cooperative game theory.

There are examples of preference profiles and sizes $k$ where no stable committee exists \citep[Thm.~4]{jiang2020approximately}. However, \citet{cheng2020group} proved that there always exists a probability distribution over committees which satisfies a probabilistic generalization of the stability property.

\begin{definition}
	A distribution $\XX \in \Delta(\power_k(A))$ over committees $X$ of size $k$ is a \emph{stable lottery} if for all alternatives $a^* \in A$, we have 
	\[ \E_{X \sim \XX} \left[ V(a^*,X) \right] < \frac{n}{k}. \] 
\end{definition}

To be self-contained, we include a short proof of existence, following the simplified treatment due to \citet[Lem.~4]{jiang2020approximately}.

\begin{theorem}
[\citealp{cheng2020group}]
	For every preference profile $\profile$ and for every $k$, there exists a stable lottery.
\end{theorem}
\begin{proof}
Let $\profile$ be a preference profile. We view our task as proving the following bound:
\[
\min_{\XX \in \Delta(\power_k(A))}
\:
\max_{a^* \in A}
\quad
\E_{X \sim \XX} [V(a^*,X)] 
<
\frac{n}{k}.
\]
If the bound holds, then an $\XX$ that solves the minimization problem is a stable lottery. We can view the expression on the left-hand side as a zero-sum game, where one player chooses a distribution and the adversary responds with an alternative. By the minimax theorem, it suffices to show that
\[
\max_{\yy \in \Delta(A)}
\:
\min_{\XX \in \Delta(\power_k(A))}
\:
\E_{X \sim \XX, a^*\sim\yy} [V(a^*,X)] 
<
\frac{n}{k}.
\]
Let $\yy \in \Delta(A)$. Define a distribution $\XX$ over committees by the following process. Draw $k$ alternatives $a_1, \dots, a_k$ from the distribution $\yy$ independently and with replacement. Let $X$ be the random set of alternatives thus selected, if necessary filled up with arbitrary additional alternatives until $|X| = k$. Now note that for every agent $i \in N$, the probability $\Pr_{a^* \sim \yy, X \sim \XX}[a^* \succ_i X]$ is at most the probability that $a^*$ is the strictly most-preferred among the at most $k+1$ alternatives $a^*, a_1, \dots, a_k$ which are drawn i.i.d. Hence by symmetry $\Pr_{a^* \sim \yy, X \sim \XX}[a^* \succ_i X] \le 1/(k+1) < 1/k$. 

Summing up over all $i \in N$, it follows that 
$\E_{X \sim \XX, a^*\sim\yy} [V(a^*,X)] 
<
\frac{n}{k}$, as desired.
\end{proof}

\citet{cheng2020group} prove that a stable lottery can be found in (expected) polynomial time using the multiplicative weights update algorithm for zero-sum games. That algorithm finds a solution whose value is $\epsilon$-close to the optimum value. But the existence proof above in fact established that the value of the game is at most $n/(k+1)$, when all we need is a solution with value less than $n/k$. Thus, we can run the algorithm with $\epsilon = \frac{1}{2} \cdot (\frac{n}{k} - \frac{n}{k+1})$ and obtain an exactly stable lottery in expected polynomial time.

\subsection{The Stable Lottery Rule}

We propose a voting rule based on stable lotteries for committees of size $k = \sqrt{m}$. Like the previously proposed harmonic rule, our rule spreads half of the probability mass uniformly over all alternatives.\footnote{Instead of $\nicefrac12$, one can use any other constant fraction (such as $0.0001$) without changing the main conclusion of \Cref{thm:distortion_ub} that the rule has distortion $O(\sqrt{m})$, though its distortion will be worse by a constant factor. One can also shift probability from a Pareto-dominated alternative to a dominating alternative without worsening distortion.}
It then assigns the remaining probability mass to alternatives in proportion to the probability that they are included in the committee selected by the stable lottery.

\begin{definition}[Stable Lottery Rule, $\SLR$]
	Let $\XX$ be a stable lottery over committees $X$ of size $k = \sqrt{m}$.
	The Stable Lottery Rule $\SLR$ works as follows: With probability $1/2$, sample a committee $X \sim \XX$ and choose an alternative uniformly at random from $X$, and with probability $1/2$, choose an alternative uniformly at random from $A$. Therefore, each alternative $a \in A$ will be selected with probability 
	$\xx(a) = \frac{1}{2\sqrt{m}} \cdot \Pr_{X\sim\XX}[a \in X] + \frac{1}{2m}$.
\end{definition}

Our first main result states that $\SLR$ achieves distortion $\Theta(\sqrt{m})$ on the class of balanced utility functions, and hence also for unit-sum, unit-range, and approval utilities. 

\begin{theorem}
	\label{thm:distortion_ub}
	On the utility class $\utilityclass^{\balanced}$, the Stable Lottery Rule achieves $O(\sqrt{m})$ distortion: 
	\[\D_m(\SLR,\utilityclass^{\balanced}) = O(\sqrt{m}).\]
\end{theorem}

\begin{proof}
	Let $\utilities$ be a utility profile consistent with a profile $\profile$, with $u_i \in \utilityclass^{\balanced}$ for all $i \in N$. 	
	We begin the proof by making the following observation. Let $X$ be a committee, and let $a^* \in A$ be a distinguished alternative. Write $u_i(X) = \sum_{a\in X} u_i(a)$ and $\SW(X,\utilities) = \sum_{i\in N} u_i(X)$. Then, 
	\begin{align}
		\textstyle
		\SW(X,\utilities) 
		&\ge 
		\textstyle
		\SW(a^*,\utilities) - V(a^*, X). \label{eq:claim}
	\end{align}
	Indeed, for every agent $i$ such that $a^* \succ_i X$, we have $u_i(X) \ge 0 \ge u_i(a^*) - \max_{a\in A} u_i(a) \ge u_i(a^*) - 1$ because $u_i \in \utilityclass^{\balanced}$, and for every agent $i$ such that $a^* \not\succ_i X$, there exists some alternative $a \in X$ such that $a \succ_i a^*$, so $u_i(X) \ge u_i(a) \ge u_i(a^*)$. \Cref{eq:claim} follows by summing these inequalities over all $i\in N$, noting that the number of agents of the first type is $V(a^*, X)$.
	
	Let $\xx = \SLR(\profile)$ be the distribution selected by the Stable Lottery Rule, and let $\XX$ be the underlying stable lottery over committees of size $\sqrt{m}$.
	Let us write $\xx = \frac12 \xx_1 + \frac12 \xx_2$, where $\xx_1$ is the part of $\xx$ based on the stable lottery and $\xx_2$ is the uniform distribution over $A$.
	Thus, $x_1(a) = \frac{1}{\sqrt{m}} \cdot \Pr_{X\sim\XX}[a \in X]$ and $x_2(a) = 1/m$ for all $a \in A$.
	
	Note that for all $i\in N$, we have $u_i(\xx_2) \ge \frac1m \sum_{a\in A} u_i(a) \ge \frac1m$ because $u_i \in \utilityclass^{\balanced}$. Hence $\SW(\xx_2,\utilities) \ge \frac nm$ and so $\frac mn \cdot \SW(\xx_2,\utilities) \ge 1$.
Now fix an arbitrary $a^* \in A$. Then,
	\begingroup
	\addtolength{\jot}{1mm} \begin{align*}
		\sqrt{m} \cdot \SW(\xx_1,\utilities)
		&= 
		\textstyle
		\sqrt{m} \cdot \sum_{a\in A} \frac{1}{\sqrt{m}} \Pr_{X\sim\XX}[a \in X] \cdot \SW(a,\utilities)
		\\
		&= \textstyle
		\E_{X \sim \XX} [\sum_{a \in X} \SW(a,\utilities)] \\
&= \textstyle
		\E_{X \sim \XX}[\SW(X,\utilities)] \\
		&\ge \textstyle
		\E_{X \sim \XX}[\SW(a^*,\utilities) - V(a^*, X)] \tag{by equation \eqref{eq:claim}} \\
		&= \textstyle
		\SW(a^*,\utilities) - \E_{X \sim \XX}[V(a^*, X)] \tag{linearity of expectation} \\
		&\ge \textstyle
		\SW(a^*,\utilities) - \tfrac{n}{\sqrt{m}}\tag{stability of $\XX$} \\
		&\ge \textstyle
		\SW(a^*,\utilities) - \frac{n}{k}  \cdot \frac{m}{n} \cdot \SW(\xx_2,\utilities) \tag{since $\frac mn \cdot \SW(\xx_2,\utilities) \ge 1$} \\
		&\ge \textstyle
		\SW(a^*,\utilities) - \frac{n}{\sqrt{m}}  \cdot \frac{m}{n} \cdot \SW(\xx_2,\utilities) \tag{since $\frac mn \cdot \SW(\xx_2,\utilities) \ge 1$} \\
		&= \SW(a^*,\utilities) - \sqrt{m} \cdot \SW(\xx_2,\utilities).
	\end{align*}
	\endgroup
	Hence,
	\[
	\SW(\xx,\utilities) =
	\tfrac12 \SW(\xx_1,\utilities) + \tfrac12\SW(\xx_2,\utilities)
	\ge \frac{\SW(a^*,\utilities)}{2\sqrt{m}} \quad \text{for all $a^* \in A$.}
	\]
	Therefore,
	\[
	\D(\xx,\profile,\utilityclass^{\balanced})
	\le
	\max_{a^* \in A} \frac{\SW(a^*,\utilities)}{\frac{\SW(a^*,\utilities)}{2\sqrt{m}}} = 2 \sqrt{m}.
	\]
	Since the above holds for all preference profiles $\profile$, we have that
	\[
	\D_m(\SLR,\utilityclass^{\balanced}) = \max_{\profile} \D(\xx,\profile,\utilityclass^{\balanced}) \le 2 \sqrt{m} = O(\sqrt{m}).
	\qedhere
	\]
\end{proof}
Following the proof that a stable lottery always exists \citep{cheng2020group}, \citet{jiang2020approximately} considered approximately stable (deterministic) committees. They proved that a 16-stable committee always exists (this proof being much more complicated than the existence of a stable lottery) but that a 2-stable committee may fail to exist. In \Cref{app:scr}, for $c \ge 1$, we adapt the Stable Lottery Rule to define the rule $\cSCR$ that uses a $c$-stable committee instead. The proof of \Cref{thm:distortion_ub} can straightforwardly be adapted to show that $\cSCR$ has distortion at most $O(c \cdot\sqrt{m})$.

In contrast to the $O(\sqrt{m})$ distortion of the Stable Lottery Rule, the Harmonic Rule $\HR$ achieves worse distortion for both unit-sum and, especially, unit-range utilities.
\begin{theorem}
	The distortion of the Harmonic Rule is $\D_m(\HR,\utilityclass^{\unitsum}) = \Theta(\sqrt{m \log m})$ for unit-sum utilities and $\D_m(\HR,\utilityclass^{\unitrange}) = \Theta(m^{2/3} \log^{1/3} m)$ for unit-range utilities.
\end{theorem}

For unit-sum utilities, the upper bound is due to \citet{BCHL+15} and the lower bound follows from the work of \citet{bhaskar2018truthful}, though we include an explicit lower bound example in \Cref{app:harmonic-distortion}. The analysis of the distortion of $\HR$ for unit-range utilities is new. The polynomial increase in the distortion of $\HR$ compared to that of $\SLR$ can be explained by noting that $\HR$ is strategyproof, and for unit-range utilities, \citet{filos2014truthfulrange} prove that any strategyproof rule has distortion $\Omega(m^{2/3})$, meaning that $\HR$ still has close to the best distortion achievable via strategyproof rules. We give proofs of these results in \Cref{app:harmonic-distortion}. 

\subsection{Lower Bounds}
\label{sec:distortion-lb}

\citet{BCHL+15} prove that the distortion of every voting rule for the class $\utilityclass^{\unitsum}$ of unit-sum utilities is $\frac12\sqrt{m} = \Omega(\sqrt{m})$, showing that $\SLR$ is optimal on this class, up to at most a constant factor of 4.\footnote{\citet{BCHL+15} prove a lower bound of $\sqrt{m}/3$, but a careful look at their analysis shows that it actually yields a lower bound of $\sqrt{m}/2$.}
Here, we present a lower bound for the class of approval utility functions, which also applies to the larger class of unit-range utilities. This bound implies that $\SLR$ achieves asymptotically optimal distortion on both of these utility classes.

\begin{theorem}
	\label{thm:LB_approval}
	For any voting rule $f$, 
	$\D_m(f,\utilityclass^{\approval}) = \Omega(\sqrt{m})$ and $\D_m(f,\utilityclass^{\unitrange}) = \Omega(\sqrt{m})$. 
\end{theorem}
\begin{proof}
	Assume $\sqrt{n}$ is a positive integer, and let $m = n + \sqrt{n}$. Each agent $i$ ranks alternative $a_i$ first, alternative $a_{n + \lceil i / \sqrt{n} \rceil}$ second, and the remaining alternatives in an arbitrary order. Note that this naturally divides the agents into $\sqrt{n}$ groups, $N_1, \dots, N_{\sqrt{n}}$, where, for $r \in [\sqrt{n}]$, $N_r$ denotes the group of agents who rank alternative $a_{n+r}$ second. 
	
	Let $f$ be a voting rule, and let $\xx$ be the distribution selected by $f$ on this profile. By the pigeonhole principle, there must exist one index $r \in [\sqrt{n}]$ such that $\xx(a_{n+r}) \le 1/\sqrt{n}$. Without loss of generality, assume that $\xx(a_{n+1}) \le 1/\sqrt{n}$. Consider the approval utility profile $\utilities$ under which all agents in $N_1$ approve their top two alternatives (i.e., their top choice and $a_{n+1}$), and all other agents approve only their top alternative.
Then, $\SW(a_{n+1},\utilities) = \sqrt{n}$ whereas $\SW(a',\utilities) = 1$ for every alternative $a' \in A \setminus \{a_{n+1}\}$. Therefore, we have	
	\[
	\D_m(f,\utilityclass^{\approval}) 
	\ge \frac{\SW(a_{n+1},\utilities)}{\SW(\xx,\utilities)} 
	\ge \frac{\sqrt{n}}{\big(1 - \frac{1}{\sqrt{n}}\big) \cdot 1 + \frac{1}{\sqrt{n}} \cdot \sqrt{n}} 
	\ge \frac{\sqrt{n}}{2} 
	= \Omega\left( \sqrt{m} \right),
	\]
where the final transition holds due to $m = n+\sqrt{n}$. Because $\utilityclass^{\approval} \subseteq \utilityclass^{\unitrange}$, we also have $\D_m(f,\utilityclass^{\unitrange}) = \Omega(\sqrt{m})$. 
\end{proof}

\section{Proportional Fairness}
\label{sec:fairness}
In this section, we turn our attention to proportional fairness (see \Cref{def:pf}). As we noted in \Cref{sec:prelim-pf}, the proportional fairness objective is scale invariant, and thus $\D^{\PF}_m(f,\utilityclass^{\all}) = \D^{\PF}_m(f,\utilityclass^{\unitsum}) = \D^{\PF}_m(f,\utilityclass^{\unitrange})$ for all voting rules $f$. We will just consider   $\utilityclass^{\all}$ throughout this section, and thus suppress the utility class $\utilityclass$ from our notation.

\subsection{Upper Bounds}

A natural question at this point is whether the stable-lottery-based approach from the previous section, which provides optimal distortion, also works for proportional fairness. In \Cref{app:scr}, we present a close cousin of our stable lottery rule, namely the \emph{stable committee rule} ($\SCR$), which uses an approximately stable \emph{deterministic} committee in place of an exactly stable \emph{lottery} over committees; such committees with constant approximations are guaranteed to exist due to the recent work of \citet{jiang2020approximately}. In \Cref{app:scr}, we show that this rule is $O(\sqrt{m})$-proportionally fair. This raises the obvious question of whether it is possible to do better. Surprisingly, we show that it is! Using the minimax theorem, we are able to show that there exists an $O(\log m)$-proportionally fair voting rule. We later show this upper bound to be tight. In \Cref{sec:fairness-computation}, we use the projected subgradient descent algorithm to turn this non-constructive argument into an efficient algorithm.

We begin with a useful lemma that simplifies the analysis of the proportional fairness of a given distribution $\xx$. Let us write $h_i(a) = \{a' \in A : a' \succeq_i a\}$ for the set of alternatives that agent $i$ ranks weakly above $a$, and for a distribution $\xx$, let $\xx(h_i(a)) = \sum_{a' \in h_i(a)} \xx(a')$ be the total weight that $\xx$ places on these alternatives.

\begin{lemma}
	\label{lem:pf-approval-simplification}
	Given a preference profile $\profile$ and a distribution $\xx$, we have
	\begin{equation}
		\label{eq:simplified-pf-distortion}
		\D^{\PF}(\xx, \profile) = 
		\max_{a \in \candids} \frac{1}{n} \sum_{i \in \voters} \frac{1}{\xx(h_i(a))},
	\end{equation}
	and $\D^{\PF}(\xx, \profile)$ is convex in $\xx$.
\end{lemma}

\begin{proof}
Recall from \Cref{sec:prelim-pf} that
\[
\D^{\PF}(\xx,\profile) = \sup_{\utilities \in (\utilityclass^{\all})^n : \utilities \triangleright \profile}\, \max_{a \in A} \frac{1}{n} \sum_{i \in N} \frac{u_i(a)}{u_i(\xx)}.
\]

Fix any $a \in A$. Note that we can take the worst case over the utility function $u_i$ of each agent $i$ separately as its contribution to the above expression, for any fixed $\xx$ and $a$, is independent of that of the other utility functions. 

Thus, it is sufficient to show that $\sup_{u_i \in \utilityclass^{\all} : u_i \triangleright \pref_i} u_i(a)/u_i(\xx) = 1/\xx(h_i(a))$. This follows from the simple observation that $u_i(a') \ge u_i(a)$ for all $a' \in h_i(a)$, which implies $u_i(\xx) \ge \xx(h_i(a)) \cdot u_i(a)$, i.e., $u_i(a)/u_i(\xx) \le 1/\xx(h_i(a))$, and noting that this upper bound is achieved by setting, for example, $u_i(a') = 1$ for all $a' \in h_i(a)$ and $u_i(a') = 0$ for all $a' \in A \setminus h_i(a)$. 

Convexity in $\xx$ follows because the function $g(z) = 1/z$ is a convex function, and taking the sum and maximum of a collection of convex functions yields a convex function. 
\end{proof}

Note that the last line of the proof shows that the worst case for proportional fairness is achieved at an approval utility profile. Hence, $\D^{\PF}_m(f,\utilityclass^{\all}) = \D^{\PF}_m(f,\utilityclass^{\approval})$ for all voting rules $f$.
Using this simplification, we can now derive an upper bound on the optimal proportional fairness.

\begin{theorem}
	\label{thm:UB-pf-logm}
There exists a voting rule $f$ with $\D^{\PF}_m(f) \le 2(1 + \ln (2m))$.
\end{theorem}

\begin{proof}
We consider the instance-optimal voting rule which, given a preference profile, selects a distribution $\xx$ that is $\alpha$-proportionally fair for the smallest $\alpha$. We interpret this distribution as the outcome of a (two-player) zero-sum game and $\alpha$ as the value of that game. We then bound this value in a dual game obtained by applying the minimax theorem.

\paragraph{Formulation as a zero-sum game.} 
Let $\profile$ be a preference profile.
Let $\D^{\PF}(\profile) = \min_{\xx \in \Delta(\candids)} \D^{\PF}(\xx, \profile)$. 
\Cref{lem:pf-approval-simplification} implies that 
\[
\D^{\PF}(\profile) = \min_{\xx \in \Delta(\candids)} \, \max_{a \in \candids} \, \frac{1}{n} \sum_{i \in \voters} \frac{1}{\xx(h_i(a))}.
\] 
Hence, $\D^{\PF}(\profile)$ can be viewed as the outcome of a zero-sum game.
The set of \emph{pure} strategies for the first player (or just the \emph{player}) is $\Delta(\candids)$, i.e., the player may choose a distribution over alternatives. In response, the second player (the \emph{adversary}) can choose a single alternative $a \in \candids$. Then, for a pair of strategies $(\xx, a) \in \Delta(\candids) \times \candids$, the payoff to the adversary, which is equal to the negative payoff of the player, is defined as
\[
\payoff(\xx, a) = \frac{1}{n} \sum_{i \in N} \frac{1}{\xx(h_i(a))}.
\]
With this notation, we have
\[
\D^{\PF}(\profile) = \min_{\xx \in \Delta(\candids)} \, \max_{a \in \candids} \, \payoff(\xx, a).
\]

Suppose we allow the adversary to choose a \emph{mixed} strategy, i.e., a distribution over alternatives $\zz \in \Delta(\candids)$. Define the expected payoff of the pair $(\xx, \zz)$ of strategies to be
$
\E_{a \sim \zz} [\payoff(\xx, a)].
$
Because this objective is linear in $\zz$, there is always a pure best response for the adversary (selecting a single alternative $a \in A$). Thus, allowing the adversary to choose a mixed strategy does not change the value of the game. Hence
\[
	\D^{\PF}(\profile) = \min_{\xx \in \Delta(\candids)}  \max_{\zz \in \Delta(\candids)} \E_{a \sim \zz} [\payoff(\xx, a)].
\]
Now note that $\E_{a \sim \zz} [\payoff(\xx, a)]$ is convex in $\xx$ (\Cref{lem:pf-approval-simplification}) and linear (and hence concave) in $\zz$. Therefore, by the minimax theorem (\Cref{thm:minimax}), we have
\[
\D^{\PF}(\profile) = \max_{\zz \in \Delta(\candids)}  \min_{\xx \in \Delta(\candids)}  \E_{a \sim \zz} [\payoff(\xx, a)].
\]
We call this game the \emph{dual game}.

\paragraph{Bounding the value of the dual game.}
	
	In the dual game, for a given strategy $\zz$ of the adversary, suppose the player responds with the strategy $\overline\xx$ with $\overline{x}(a) = \frac{1}{2} z(a) + \frac{1}{2m}$ for all $a \in A$ (which is not necessarily a best response). Thus, with probability $\frac{1}{2}$ the player selects according to $\zz$, and with probability $\frac{1}{2}$, the player selects an alternative uniformly at random. Note that the value of the dual game when the player plays $\overline\xx$ is an upper bound on the true value of the dual game. Now, we have
	\begin{align}
		\D^{\PF}(\profile)
		&=
		\max_{\zz \in \Delta(\candids)}  \min_{\xx \in \Delta(\candids)}  \E_{a \sim \zz} [\payoff(\xx, a)]
		&
		\nonumber
		\\
		&
		\le
		\max_{\zz \in \Delta(\candids)}  \E_{a \sim \zz} [\payoff(\overline\xx, a)]
		\tag{first player responds with $\overline\xx$}
		\\
		&=
		\max_{\zz \in \Delta(\candids)} \frac{1}{n} \sum_{i \in N}   \E_{a \sim \zz}  \left[  \frac{1}{\overline{x}(h_i(a))}\right]
		\tag{linearity of expectation}
		\\
		&\le
		\max_{\zz \in \Delta(\candids)} \, \max_{i \in N} \,  \E_{a \sim \zz} \left[ \frac{1}{\overline{x}(h_i(a))}\right].
		\label{eq:dual-value-upper-bound1}
	\end{align}
	The last term is maximized at some distribution $\zz$ and some agent $i$ with preference ranking $\pref_i$.
	Without loss of generality, suppose $\pref_i = a_1 \succ_i a_2 \succ_i \dots \succ_i a_m$. Write $T_j = \sum_{\ell = 1}^j \overline{x}(a_\ell) = \overline{x}(h_i(a_j))$ and $T_0 = 0$. Then \eqref{eq:dual-value-upper-bound1} is equal to	
	\begin{align*}
		\sum_{j \in [m]} z(a_j) \cdot \frac{1}{\overline{x}(h_i(a_j))}
		<
		\sum_{j \in [m]} \frac{2 \cdot \overline{x}(a_j)}{\overline{x}(h_i(a_j))}
		=
		2 \sum_{j \in [m]} \frac{T_{j} - T_{j - 1}}{T_j}
		=
		2 \sum_{j \in [m]} \left(1 - \frac{T_{j - 1}}{T_j}\right).
	\end{align*}
	Using the fact that $1 - x \le -\ln(x)$, 
	\begin{align*}
		\sum_{j \in [m]} \left(1 - \frac{T_{j - 1}}{T_j}\right) \le 
		1 + \sum_{j = 2}^m  \left(\ln(T_j) - \ln(T_{j - 1})\right)
		= 1 + \ln(T_m) - \ln(T_1)  \le 1 + \ln(2m),
	\end{align*}
	where the last inequality holds due to $T_m = 1$ and $T_1 = \xx(a_1) \ge \frac{1}{2m}$. It follows that $\D^{\PF}(\profile) \le 2(1 + \ln(2m))$, as desired.
\end{proof}

This upper bound on proportional fairness immediately implies upper bounds on the universal $\alpha$-core and distortion with respect to Nash welfare, using \Cref{prop:pf-core,prop:pf-nash}.
\begin{corollary}
	\label{cor:UB-core-nash-logm}
	Let $\alpha = 2(1 + \ln (2m))$. 	There exists a voting rule $f$ which is in the universal $\alpha$-core and whose distortion with respect to Nash welfare is $\D_m^{\NW}(f) \le \alpha$.
\end{corollary}

What about strategyproof rules? In the context of distortion with respect to utilitarian social welfare, we saw that strategyproof rules (in particular, the harmonic rule $\HR$) can provide a distortion that is only a logarithmic factor worse than the optimum. In the context of proportional fairness, however, strategyproofness comes at a much larger cost. Indeed, strategyproof rules must be exponentially worse than the optimum: strategyproof rules cannot be better than $\Omega(\sqrt{m})$-proportionally fair, when the optimum is $O(\log m)$-proportionally fairness. This lower bound is again almost attained by the harmonic rule $\HR$, which is $\Theta(\sqrt{m \log m})$-proportionally fair.
\begin{theorem}
	The harmonic rule $\HR$ is $\Theta(\sqrt{m \log m})$-proportionally fair.
\end{theorem}
\begin{theorem}
	If $f$ is an $\alpha$-proportionally fair voting rule that is also strategyproof, then $\alpha = \Omega(\sqrt{m})$.
\end{theorem}
We provide proofs of these results in \Cref{sec:hr-pf} and \Cref{sec:truthful-pf}, respectively.

\subsection{Lower Bounds}

Next, we give a lower bound that matches our upper bound up to a constant factor. We thank an anonymous reviewer for suggesting this lower bound construction.

\begin{theorem}
	\label{thm:LB-pf-2}
	Every voting rule $f$ has distortion at least $\frac12 \log_2 m$ with respect to the Nash welfare and with respect to proportional fairness. Furthermore, if $f$ is in the universal $\alpha$-core, then $\alpha \ge \frac18 \log_2 m$.
\end{theorem}

\begin{proof}
First, we derive the lower bound on the distortion with respect to the Nash welfare and proportional fairness. Then, we show that the same construction provides the desired core lower bound as well.

\paragraph{Nash welfare and proportional fairness lower bound.} We show that there exists a preference profile $\profile$ for which $\D^{\NW}(\xx,\profile) \ge \Omega(\log m)$ for every distribution $\xx \in \Delta(A)$.

For simplicity, we assume $n = 2^{k - 1}$ and $m = 2^k - 1$, but the proof works when $n$ is a multiple of $2^{k - 1}$. Take the preference profile $\profile$, in which the $k$th vote of all agents is alternative $a_1$, the $(k - 1)$th vote of all agents is evenly divided between $a_2$ and $a_3$, the $(k - 2)$th votes are equally divided between $a_4, \ldots, a_7$, and similarly for $\ell \in [k]$ the $\ell$th vote of all agents is divided equally among alternatives $a_{2^{k - \ell}}, \ldots, a_{2^{k  - \ell + 1} - 1}$ each with $2^{\ell - 1}$  votes at rank $\ell$. We fill the bottom $m - k$ votes of all agents arbitrarily. A construction for when $k = 4$ is depicted in \Cref{tbl:example-lowerbound-nash}.

As usual, we can think about distortion as a zero-sum game. For every utility profile $\utilities$ and distribution $\yy$, we have $\D^{\NW}(\xx) \ge \NW(\yy, \utilities) / \NW(\xx, \utilities)$, and we can think of $\utilities$ and $\yy$ as being chosen by an \emph{adversary} that maximizes the right-hand quantity.
To obtain a lower bound, we weaken the adversary and assume the realized utility profile is one of the $k$ utility profiles described as follows. For $\ell \in [k]$, let $\utilities_{\ell}$ be the utility profile where each agent has a utility of $1$ for their top $\ell$ votes and a utility of $0$ for their bottom $m - \ell$ votes. On $\utilities_{\ell}$, suppose the adversary selects the distribution $\yy_\ell$ that is the uniform distribution over the alternatives appearing on the $\ell$th rank, i.e., $y_\ell(a_{j}) = 2^{-(k - \ell)}$ for $j \in [2^{k - \ell}, 2^{k - \ell + 1} - 1]$. Then, for all agents $i \in [n]$, $u_i(\yy_\ell, \utilities_{\ell}) = 2^{-(k - \ell)}$. As a result, $\NW(\yy_{\ell}, \utilities_{\ell}) = 2^{-(k - \ell)}$. Next, we show that every distribution $\xx \in \Delta(A)$ incurs a distortion of at least $k / 2s$ with respect to one of the $\utilities_{\ell}$'s.

By the inequality of arithmetic and geometric means, we have $\NW(\xx) \le \frac{1}{n} \sum_{i \in [n]} u_{i}(\xx)$ for all utility profiles.
For all $\ell \in [k]$,
\begin{align*}
\NW(\xx, \utilities_{\ell}) \le \frac{1}{n} \sum_{i \in [n]} u_{i}(\xx, \utilities_{\ell})
&=
\frac{1}{n} \sum_{i \in [n]} \sum_{j \in [m]} \xx(a_j) \cdot \mathbb{1}[\rank_i(a_j) \le \ell]
\\
&=
\frac{1}{n} \sum_{j \in [m]}  \; \xx(a_j) \cdot \left|\{i \in [n] \mid \rank_i(a_j) \le \ell\}\right|,
\\
&= \frac{1}{n} \sum\nolimits_{\ell' = 1}^{\ell} \; \sum\nolimits_{j = 2^{k - \ell'}}^{2^{k - \ell' + 1} - 1} \; \xx(a_j) \cdot 2^{\ell' - 1},
\end{align*}
where in the last transition we regrouped the summation based on the positions an alternative takes among the top $\ell$ votes of the preference profile. Denote the total probability mass on the alternatives that appear on the $\ell$th rank by $p_{\ell} = \sum\nolimits_{j = 2^{k - \ell}}^{2^{k - \ell + 1} - 1} \; \xx(a_j)$. Then, the above is equal to
\[
\frac{1}{n} \sum_{i \in [n]} u_{i}(\xx, \utilities_{\ell}) = \frac{1}{n} \sum\nolimits_{\ell' = 1}^{\ell} p_{\ell'} \cdot 2^{\ell' - 1} = \sum\nolimits_{\ell' = 1}^{\ell} p_{\ell'} \cdot 2^{\ell' - k},
\]
and distortion of $\xx$ w.r.t. the Nash welfare is at least
\begin{equation}
\label{eq:nash-lb-am-gm}
\D^{\NW}(\xx, \utilities_{\ell}) \ge \frac{\NW(\yy_{\ell}, \utilities_{\ell})}{\NW(\xx_{\ell}, \utilities_{\ell})} \ge \frac{2^{-(k - \ell)}}{\sum\nolimits_{\ell' = 1}^{\ell} p_{\ell'} \cdot 2^{\ell' - k}} = \frac{1}{\sum\nolimits_{\ell' = 1}^{\ell} p_{\ell'} \cdot 2^{\ell' - \ell}}.
\end{equation}
Next, using an averaging argument, we will show for at least  one value of $\ell \in [k]$, the denominator above $\sum\nolimits_{\ell' = 1}^{\ell} p_{\ell} \cdot 2^{\ell' - \ell} < \frac{2}{k}$ and hence $\D^{\NW}(\xx) \ge \D^{\NW}(\xx, \utilities_{\ell}) > \frac{k}{2}$. It follows from
\begin{align}
\min_{\ell \in [k]} \, \sum\nolimits_{\ell' = 1}^{\ell} p_{\ell'} \cdot 2^{\ell' - \ell}
&\le
\frac{1}{k}  \sum\nolimits_{\ell = 1}^k  \, \sum\nolimits_{\ell' = 1}^{\ell} p_{\ell'} \cdot 2^{\ell' - \ell}
\nonumber
\\
&= \frac{1}{k} \sum_{\ell' = 1}^{k} p_{\ell'} \sum_{\ell = \ell'}^k 2^{\ell' - \ell}
\nonumber
\\
&= \frac{1}{k} \sum_{\ell' = 1}^{k} p_{\ell'} \cdot (2 - 2^{\ell' - k}) < \frac{1}{k} \sum_{\ell' = 1}^{k} p_{\ell'} \cdot 2 = \frac{2}{k}.
\label{eq:nash-lb-logm}
\end{align}
Hence, $\D^{\NW}(\xx) > \frac{k}{2} \ge \frac{\log_2 m}{2}$ for all $\xx \in \Delta(A)$. Using \Cref{prop:pf-nash}, we also have $\D^{\PF}(\xx) > \frac{\log_2 m}{2}$.

\begin{table}[t]
	\centering
	\begin{tabular}{cccccccc}
		\toprule
		$i_1$ & $i_2$ & $i_3$  & $i_4$  & $i_5$  & $i_6$  & $i_7$  & $i_8$ \\
		\midrule
		$a_{8}$ & $a_{9}$ & $a_{10}$ & $a_{11}$ & $a_{12}$ & $a_{13}$ & $a_{14}$ & $a_{15}$ \\
		$a_{4}$ & $a_{4}$ & $a_{5}$ & $a_{5}$ & $a_{6}$ & $a_{6}$ & $a_{7}$ & $a_{7}$ \\
		$a_{2}$ & $a_{2}$ & $a_{2}$ & $a_{2}$ & $a_{3}$ & $a_{3}$ & $a_{3}$ & $a_{3}$\\
		$a_{1}$ & $a_{1}$ & $a_{1}$ & $a_{1}$ & $a_{1}$ & $a_{1}$ & $a_{1}$ & $a_{1}$ \\
		$\vdots$ & $\vdots$ & $\vdots$ & $\vdots$ & $\vdots$ & $\vdots$ & $\vdots$ & $\vdots$ \\
		\bottomrule
	\end{tabular}
	\caption{Lower bound instance for distortion with the Nash welfare with $m = 2^4 - 1$ alternatives.}
	\label{tbl:example-lowerbound-nash}
\end{table}

\paragraph{Core lower bound.} Under  $\utilities_{\ell}$ and $\yy_{\ell}$,  all agents  have the same utility $u_i(\yy_{\ell}, \utilities_{\ell}) = 2^{-(k - \ell)} = \gamma$. From the above analysis, we have
\[
\NW(\yy_{\ell}, \utilities_{\ell}) = \gamma > \frac{k}{2} \cdot \Big( \frac{1}{n} \cdot \sum_{i \in [n]} u_i(\xx, \utilities_{\ell}) \Big).
\] 
Then, there must exist at least $n/2$ agents $i \in \voters' \subseteq \voters$ with $u_i(\xx, \utilities_{\ell}) < \frac{4}{k} \cdot \gamma$. (Otherwise the RHS would be at least $\frac{k}{2} \cdot \frac{1}{n} \cdot (\frac{n}{2} \cdot \frac{4}{k} \cdot \gamma)  \ge \gamma$.) This violates the $(\nicefrac{k}{8})$-core since for all $i \in \voters'$,
\[
\frac{|\voters'|}{|\voters|} \cdot  u_i(\yy_{\ell}, \utilities_{\ell}) \ge \frac{1}{2} \cdot \gamma = \frac{k}{8} \cdot \left( \frac{4}{k} \cdot \gamma \right)
\ge \frac{k}{8} \cdot u_i(\xx, \utilities_{\ell}).
\qedhere
\]
\end{proof}

\subsection{Computation}\label{sec:fairness-computation}
\Cref{lem:pf-approval-simplification} gives a simple formula for calculating the value $\D^{\PF}(\xx, \profile)$ of a given distribution $\xx$.
Now, we turn to the computational problem of finding a distribution $\xx$ with the lowest possible distortion with respect to proportional fairness, for a given preference profile $\profile$. We show that this problem can be (approximately) solved in polynomial time. Our argument depends on the convexity of $\D^{\PF}(\xx, \profile)$ in $\xx$ (\Cref{lem:pf-approval-simplification}), which allows us to use convex optimization methods (in particular, the projected subgradient descent algorithm). For definitions of a subgradient and the subdifferential $\partial f$ of a convex function $f$, we refer the reader to the books of \citet{nesterov2003introductory} and \citet{vishnoi2021algorithms}.

\begin{theorem}
	[\Citet{nesterov2003introductory}, Chapter~3.2, \Citet{vishnoi2021algorithms}, Theorem~7.1]
	\label{thm:projected-sgd}
	Let $f$ be a convex function over a bounded closed convex set $Q$.
	There is an algorithm (based on subgradient descent) that, given
	(a)
an oracle that, given $\xx \in Q$, can return $f(\xx)$ and a subgradient $\mathbf{g} \in \partial f(\xx)$;
	(b) a number $G$ such that for all $\xx \in Q$ and subgradients $\mathbf{g} \in \partial f(\xx)$, we have $\left\| \mathbf{g} \right\|_2 \le G$; (c) an initial point $\xx^0 \in Q$; (d) a number $D$ such that $\left\|\xx^0 - \xx^*\right\|_2 \le D$ where $\xx^* = \arg\min_{\xx \in Q} f(\xx)$; and (e) some $\epsilon > 0$, outputs a sequence $\xx^0, \xx^1, \ldots, \xx^{T - 1}$ such that $\frac{1}{T} \sum_{i = 0}^{T-1} f(\xx^i) \; - \; f(\xx^*) \le \epsilon$, where $T = \left(\frac{DG}{\epsilon}\right)^2$.
\end{theorem}

We can use this algorithm to compute the optimal distribution $\xx$, as we show in the following theorem. In practice, we can solve the relevant convex optimization problem using standard solvers, for example using the \texttt{cvxpy} package.\footnote{Example code is available at \url{https://gist.github.com/DominikPeters/8fced1e221783781129e24f4ac5dce8b}.}

\begin{theorem}
	Given a preference profile $\profile$ and $\epsilon > 0$, writing $\xx^*$ for the distribution optimizing distortion with respect to proportional fairness, a distribution $\xx$ with $\D^{\PF}(\xx, \profile) \le \D^{\PF}(\xx^*, \profile)  + \varepsilon$ can be computed in $\poly(n, m, 1/\varepsilon)$ time.
\end{theorem}

\begin{proof}
	We apply the projected subgradient method described in \Cref{thm:projected-sgd} to the function $f(\xx) = \D^{\PF}(\xx, \profile) = \max_{a \in A} \payoff(\xx,a)$, where $\payoff(\xx,a) = \frac{1}{n} \sum_{i \in N} 1/\xx(h_i(a))$. We have shown $f$ to be convex in $\xx$ (\Cref{lem:pf-approval-simplification}). Note that for a given $\xx \in \Delta(\candids)$, $f(\xx)$ can be computed in $\poly(n, m)$ time. We want to minimize $f$ over $\xx \in \Delta(A)$. However, the subgradients at points close to the boundary of $\Delta(A)$ can be unbounded. We will avoid this issue by carefully restricting the domain of $f$. 

For each $a \in A$, let $p_a$ be the fraction of agents who rank $a$ as their top choice. Let $\beta = 2(1 + \ln(2m))$ be the upper bound proven in \Cref{thm:UB-pf-logm} on the optimal distortion with respect to proportional fairness. We set $Q = \{ \xx \in \Delta(\candids) : \xx(a) \ge  {p_a}/{\beta}, \forall a \in \candids\}$. 

First, we will show that an optimal distribution $\xx^* \in \argmin_{\xx \in \Delta(\candids)} f(\xx)$ lies in $Q$, ensuring that it is sufficient to optimize $f$ over $Q$. Subsequently, we will show that the norm of any subgradient of $f$ at any point in $Q$ is bounded by $\poly(n, m)$ and that such a subgradient can be computed in polynomial time, giving us conditions (a) and (b) of \Cref{thm:projected-sgd}. Then, the only missing piece left to be able to apply \Cref{thm:projected-sgd} is a starting point: we can choose any $\xx^0 \in Q$ (e.g., $\xx^0(a) = p_a$ for each $a \in A$) and use $D = \sqrt{2}$, which is an upper bound on the Euclidean distance between any two probability distributions. 

\paragraph{Optimality.} Let $\xx^* \in \argmin_{\xx \in \Delta(\candids)} f(\xx)$. Assume for a contradiction that $\xx^* \notin Q$. Thus, there exists an alternative $a \in A$ with $p_a > 0$ such that $\xx^*(a) <  p_a/\beta$. Then,
\begin{align*}
		\D^{\PF}(\xx^*, \profile) \ge  \frac{1}{n} \sum_{i \in N} \frac{1}{\xx^*(h_i(a))} \ge p_a  \cdot \frac{1}{\xx^*(a)} > \beta.
\end{align*}
	This contradicts \Cref{thm:UB-pf-logm}, where we proved that $\D^{\PF}(\xx^*, \profile) \le \beta$. Therefore, $\xx^* \in Q$.

	\paragraph{Bounding and computing subgradients.} Take any $\xx \in Q$. For each fixed $a \in \candids$, the function $\payoff(\xx, a)$ is differentiable and convex in $\xx$.
More specifically, for all $a,a' \in \candids$, we have 
	\[
	\frac{\partial}{\partial x_{a'}} \payoff(\xx, a) =  \frac{1}{n} \sum_{i \in N : a' \succeq_i a} \frac{-1}{\xx(h_i(a))^2}.
	\]
	For each $i \in N$, let $a^*_i$ be the top alternative of agent $i$. Because $p_{a^*_i} \ge 1/n$ and $\xx \in Q$, we have 
	\[
	\xx(h_i(a)) \ge \xx(a^*_i) \ge \frac{p_{a^*_i}}{\beta} \ge \frac{1}{2n(1 + \ln (2m))}.
	\]
	Therefore, $\lVert \nabla \payoff(\xx, a)  \rVert_{\infty} = O((n \ln m)^2)$ for all $a$.
	Furthermore, it is known that any gradient $\nabla \payoff(\xx, a^*)$, where $a^* \in \argmax_{a \in \candids} \payoff(\xx, a)$ is a subgradient of $\D^{\PF}(\xx, \profile) = \max_{a \in \candids} \payoff(\xx, a)$. Hence, it follows that the norm of such a subgradient is bounded by $\poly(n,m)$ and we can compute such a subgradient in $\poly(n, m)$ time. 
\end{proof}

\section{Discussion}\label{sec:discussion}

We have proved that the best distortion (with respect to the utilitarian welfare) that probabilistic voting rules can achieve with ranked preferences is $\Theta(\sqrt{m})$, resolving an open question by \citet{BCHL+15}. We have also initiated the study of the distortion of the proportional fairness objective, which focuses on fairness rather than efficiency. We proved that the worst-case distortion of this objective with ranked preferences is $\Theta(\log m)$. The same bound applies to the distortion with respect to Nash welfare. Similarly, one can also focus on distortion with respect to other welfare functions, such as the egalitarian welfare or, more generally, the $p$-mean welfare~\citep{barman2020tight,chaudhury2021fair}. For the egalitarian welfare, it is easy to see that the best distortion for both unit-sum and approval utilities is $\Theta(m)$.\footnote{The upper bound in both cases can be achieved by assigning a probability of $1/m$ to each alternative. For the lower bound, consider a profile over $m$ alternatives $a_1,\ldots,a_m$, in which the agents are partitioned into $m-1$ equal-sized groups. For $i \in [m-1]$, agents in group $i$ rank $a_i$ first, $a_m$ second, and the remaining alternatives arbitrarily. Any probabilistic voting rule must assign probability at most $1/(m-1)$ to at least one of $a_1,\ldots,a_{m-1}$, say to $a_i$. It is possible that agents in group $i$ have utility $1$ for $a_i$ and $0$ for the remaining alternatives, while agents in every other group $j$ have utility $1/2$ (resp., $1$) for $a_j$ and $a_m$ and $0$ for the remaining alternatives in the unit-sum (resp., approval) case. The egalitarian welfare achieved by the rule is at most $1/(m-1)$ in both cases due to the agents in group $i$. In contrast, the distribution that assigns probability $1/2$ to both $a_i$ and $a_m$ achieves $\Omega(1)$ egalitarian welfare in both cases, yielding an $\Omega(m)$ lower bound on the best possible distortion with respect to the egalitarian welfare in both cases.}

In \Cref{app:nash-two}, we discuss the case of $m = 2$ alternatives, which is interesting for referenda and for studying pairwise comparisons. We are able to characterize the instance-optimal voting rules for all of our objective functions. Interestingly, compared to the utilitarian objectives, the rules for the fairness objectives (proportional fairness and Nash welfare) stay closer to the 50/50 uniform distribution. The worst-case distortion for $m = 2$ turns out to be $\sqrt{2} \approx 1.41$ for Nash welfare and $1.5$ for proportional fairness. For utilitarian welfare, we find $1.5$ for unit-sum utilities and $4/3$ for unit-range utilities.
Beyond our setting, there is significant literature on studying distortion with respect to the utilitarian welfare for ballot formats other than ranked preferences~\citep{BNPS21,MPSW19,MPW20,amanatidis2021peeking,borodin2022distortion}. A natural direction for future work is to study proportional fairness and distortion with respect to other welfare functions for such ballot formats. One can also extend these ideas from single-winner selection to committee selection, where the output of a voting rule is a (randomized) subset of alternatives of a given size, and participatory budgeting, where each alternative has a cost and the output is a (randomized) subset of alternatives with total cost at most a given budget. 

Finally, centuries of research on voting theory has focused on \emph{simple} voting rules (such as plurality or Borda count) that are easy for voters to understand and satisfy appealing axiomatic properties. A significant barrier to the modern optimization-based approaches, which focus on quantitative objectives such as distortion or proportional fairness, is that they often yield rules that are difficult to understand (and sometimes difficult to compute). Significant challenges lie ahead in paving the path for increased practicability of such approaches: Can we design simple rules that perform well on these quantitative metrics? Alternatively, can we convey the intricate rules emerging from such approaches to the end users by providing simple-to-digest \emph{explanations} of either their end goal or their properties~\citep{peters2021market}? Can we reconcile these quantitative approaches with the classical axiomatic approach to find rules that achieve the best of both worlds?

\subsection*{Acknowledgements}
We thank the anonymous reviewers for their feedback that helped improve the presentation of the paper, and we thank one reviewer for suggesting the lower bound construction of \Cref{thm:LB-pf-2}.

\bibliographystyle{plainnat}

\appendix

\newpage
\section*{Appendix}

\section{Approval vs. Unit-Range Utilities}\label{app:approval-unit-range}

In this section, we show that approval utilities are the worst case for distortion among the broader class of unit-range utilities. Hence, any upper bounds derived for approval utilities (like in \Cref{thm:hr-approval} in the next section) apply to the broader class of unit-range utilities as well. The proof also implies that a distribution $\xx$ that minimizes $\D(\xx, \profile, \utilityclass^{\unitrange})$ thereby also minimizes $\D(\xx, \profile, \utilityclass^{\approval})$, which means that we can use a linear program to optimize the latter quantity (see the discussion after \Cref{exm:distortion}).

\begin{lemma}\label{lem:approval-unit-range}
	For every voting rule $f$, we have $\D_m(f,\utilityclass^{\unitrange}) = \D_m(f, \utilityclass^{\approval})$.
\end{lemma}
\begin{proof}
	As $\utilityclass^{\approval} \subseteq \utilityclass^{\unitrange}$, we trivially have $\D_m(f, \utilityclass^{\approval}) \le \D_m(f, \utilityclass^{\unitrange})$. We show that the inequality also holds in the opposite direction. We will prove a stronger argument: for every distribution $\xx$ and preference profile $\profile$, we have $\D(\xx,\profile,\utilityclass^{\unitrange}) \le \D(\xx,\profile,\utilityclass^{\approval})$. Fix any distribution $\xx$ and preference profile $\profile$.  
	
Let $\utilities \in (\utilityclass^{\unitrange})^n$ be a utility profile consistent with $\profile$ that maximizes $\D(\xx,\utilities)$, and among all such utility profiles, let it be one that minimizes the number of agents who do not have approval utilities. If $\utilities \in (\utilityclass^{\approval})^n$, then we are done. Suppose this is not the case. Fix any agent $i$ such that $u_i \notin \utilityclass^{\approval}$. 

Let $a^* \in \argmax_{a \in A} \SW(a,\utilities)$ be an alternative maximizing utilitarian welfare under $\utilities$. Then, 
\begin{align}\label{eqn:dependence-on-i}
\D(\xx,\utilities) = \frac{\SW(a^*,\utilities)}{\SW(\xx,\utilities)} = \frac{\SW(a^*,\utilities_{-i})+u_i(a^*)}{\SW(\xx,\utilities_{-i})+\sum_{a \in A} \xx(a) \cdot u_i(a)},
\end{align}
where $\utilities_{-i}$ denotes the utility profile containing the utility functions of all agents except agent $i$. 

If $a^*$ is the top alternative of agent $i$ (i.e., $\pref_i(a^*) = 1$), then by the definition of unit-range utilities, we must have $u_i(a^*) = 1$. In that case, it is easy to see that the expression in \eqref{eqn:dependence-on-i} is maximized when $u_i(a) = 0$ for all $a \in A\setminus\set{a^*}$. That is, define $\utilities^*$ such that $\utilities^*_{-i} = \utilities_{-i}$, $u^*_i(a^*) = 1$, and $u^*_i(a) = 0$ for all $a \in A\setminus\set{a^*}$. Then, $\D(\xx,\utilities^*) \ge \D(\xx,\utilities)$, which is a contradiction because $\utilities^*$ has at least one more agent with an approval utility function compared to $\utilities$. 

Now, suppose $\pref_i(a^*) \ge 2$. Denote by $a^+$ the top alternative of agent $i$ satisfying $\pref_i(a^+) = 1$. Write $\bar{h}_i(a^*) = \set{a \in A \setminus \set{a^+} : a \succeq_i a^*}$. Consider a different utility profile $\utilities'$, where $\utilities'_{-i} = \utilities_{-i}$, $u'_i(a^+) = u_i(a^+) = 1$, $u'_i(a) = u_i(a^*)$ for all $a \in \bar{h}_i(a^*)$, and $u_i(a) = 0$ for all $a \in A$ with $a^* \succ_i a$. Note that we are reducing the utility of agent $i$ for any alternative she ranks higher than $a^*$ to $u_i(a^*)$, and reducing her utility for any alternative she ranks lower than $a^*$ to $0$, without changing her utility for $a^*$. This can only (weakly) reduce the denominator in \eqref{eqn:dependence-on-i} without changing the numerator, implying that $\D(\xx,\utilities') \ge \D(\xx,\utilities)$. 

Next, notice that  
\begin{align*}
\D(\xx,\utilities') &= \frac{\SW(a^*,\utilities_{-i})+u_i(a^*)}{\SW(\xx,\utilities_{-i})+\xx(a^+) \cdot 1 +\xx(\bar{h}_i(a^*)) \cdot u_i(a^*)} \\
&\le \max\left(\frac{\SW(a^*,\utilities_{-i})+1}{\SW(\xx,\utilities_{-i})+\xx(a^+) \cdot 1 + \xx(\bar{h}_i(a^*)) \cdot 1} , \frac{\SW(a^*,\utilities_{-i})}{\SW(\xx,\utilities_{-i})+\xx(a^+) \cdot 1}\right),
\end{align*}
where the final transition holds due to the weighted mediant inequality which states that for all $t$, all weights $\mathbf{w} \in \Delta([t])$, and all positive numbers $a_1, \dots, a_t$ and $b_1, \dots, b_t$, we have $\min_{i \in [t]} \frac{a_i}{b_i} \le \frac{\sum_{i \in [t]} \mathbf{w}(i) a_i}{\sum_{i \in [t]} \mathbf{w}(i) b_i} \le \max_{i \in [t]} \frac{a_i}{b_i}$. Here, we take $t = 2$ and 
\begin{align*}
(a_1,b_1) &= (\SW(a^*,\utilities_{-i}),\SW(\xx,\utilities_{-i})+\xx(a^+)), \\
(a_2,b_2) &= (\SW(a^*,\utilities_{-i})+1,\SW(\xx,\utilities_{-i})+\xx(a^+)+\xx(\bar{h}_i(a^*)))\text{, and} \\ 
(\mathbf{w}(1),\mathbf{w}(2)) &= (1-u_i(a^*),u_i(a^*)).
\end{align*}

Hence, we can see that one of two choices --- either increasing the utility of agent $i$ for all the alternatives in $\bar{h}_i(a^*)$ to $1$ or decreasing them all to $0$ --- does not reduce the distortion. Making this choice yields another utility profile $\utilities^*$ consistent with $\profile$ such that $\D(\xx,\utilities^*) \ge \D(\xx,\utilities') \ge \D(\xx,\utilities)$, but $\utilities^*$ has at least one more agent having an approval utility function, which is a contradiction.
\end{proof}

In the proof of \Cref{lem:pf-approval-simplification}, we proved that the same conclusion holds for proportional fairness. Because proportional fairness is scale-invariant, this conclusion actually holds with respect to the class $\utilityclass^{\all}$ of all utility functions.

\begin{lemma}\label{lem:pf-approval-unit-range}
	For every voting rule $f$, we have $\D^{\PF}_m(f,\utilityclass^{\all}) = \D^{\PF}_m(f,\utilityclass^{\approval})$.
\end{lemma}

Using a slight generalization of that argument, one can also prove this result for distortion with respect to Nash welfare. 
\begin{lemma}\label{lem:nash-approval-unit-range}
	For every voting rule $f$, we have $\D^{\NW}(f,\utilityclass^{\all}) = \D^{\NW}(f,\utilityclass^{\approval})$.
\end{lemma}
\begin{proof}
We prove a stronger result: for every distribution $\xx$ and every preference profile $\profile$, we have $\D^{\NW}(\xx,\profile,\utilityclass^{\all}) = \D^{\NW}(\xx,\profile,\utilityclass^{\approval})$. Recall that 
\[
\D^{\NW}(\xx,\profile,\utilityclass^{\all}) = \sup_{\utilities \in (\utilityclass^{\all})^n : \utilities \triangleright \profile} \sup_{\yy} \left(\prod_{i \in N} \frac{u_i(\yy)}{u_i(\xx)}\right)^{1/n}.
\]

First, as in the proof of \Cref{lem:pf-approval-simplification}, we see that we can take the worst case over the utility function $u_i$ of each agent $i$ separately as its contribution to the distortion expression is independent of that of the other utility functions. Thus, it is sufficient to prove that for fixed distributions $\xx$ and $\yy$ and agent $i$, there is an approval utility function $u_i$ that maximizes $\frac{u_i(\yy)}{u_i(\xx)}$ across all utility functions consistent with $\profile$. Fix any distributions $\xx$ and $\yy$, agent $i$, and utility function $u_i : A \to \bbR_{\ge 0}$. 
	
For simplicity, label alternatives so that $\pref_i$ is  $a_1 \succ_i a_2 \succ_i \dots \succ_i a_m$, and hence $u_i(a_1) \ge \dots \ge u_i(a_m)$. Take the $m$ different approval utility functions consistent with $\pref_i$: for all $j \in [m]$, let $v_j$ be the utility function that approves alternatives $a_1$ to $a_j$. Note that $u_i$ can be written as a non-negative linear combination of the approval utilities, that is, $u_i = \sum_{j \in [m]} \alpha_j v_j$ for some $\alpha_1, \dots, \alpha_m \ge 0$. (Explicitly, we can take $\alpha_m = u_i(a_m)$ and $\alpha_j = u_i(a_j) - u_i(a_{j + 1}) \ge 0$ for each $j < m$.) 
Because the Nash welfare is scale-free, we can rescale the utility function $u_i$ and the coefficients $\alpha_j$ such that $\sum_{j \in [m]} \alpha_j = 1$.
Then, 
\[
\frac{u_i(\yy)}{u_i(\xx)} = \frac{\sum_{j \in [m]} \alpha_j v_j(\yy)}{\sum_{j \in [m]} \alpha_j v_j(\xx)}
\le \max_{j \in [m]} \frac{v_j(\yy)}{v_j(\xx)},
\]
where the final transition is due to the weighted mediant inequality which states that for all $t$, all weights $\mathbf{w} \in \Delta([t])$, and all positive numbers $a_1, \dots, a_t$ and $b_1, \dots, b_t$, we have 
\[\min_{k \in [t]} \frac{a_k}{b_k} \le \frac{\sum_{k \in [t]} \mathbf{w}(k) a_k}{\sum_{k \in [t]} \mathbf{w}(k) b_k} \le \max_{k \in [t]} \frac{a_k}{b_k}.\]
This proves that $u_i(\yy)/u_i(\xx) \le v_j(\yy)/v_j(\xx)$ for some approval utility function $v_j$, as desired.
\end{proof}

\section{The Harmonic Rule}
\label{app:harmonic}
In this section, we provide a detailed analysis of the harmonic rule $\HR$ proposed by \citet{BCHL+15}. Recall that for each $a \in A$, we write $\harm(a) := \sum_{i \in N} 1 / \pref_\voter(a)$ for its harmonic score. With probability $1/2$, $\HR$ chooses an alternative uniformly at random, and with probability $1/2$, $\HR$ chooses an alternative proportionally to its harmonic score. In other words, the rule chooses each $a \in A$ with probability $\xx(a) := \frac{1}{2m} + \frac{\harm(a)}{2 \sum_{a' \in A} \harm(a')}$. Note that $\sum_{a' \in A} \harm(a') = n H_m$, where $H_m := \sum_{i \in [m]} 1/m$ is the $m^{th}$ harmonic number, so we may rewrite $\xx(a) = \frac{1}{2m} + \frac{\harm(a)}{2 n H_m}$.

\subsection{Distortion}
\label{app:harmonic-distortion}
\citet{BCHL+15} show that the distortion of the harmonic rule for unit-sum utilities satisfies $\D_m(\HR,\utilityclass^{\unitsum}) = O(\sqrt{m \log m})$. \citet{bhaskar2018truthful} show that every strategyproof rule $f$  incurs distortion $\D_m(f,\utilityclass^{\unitsum}) = \Omega(\sqrt{m \log m})$. Since $\HR$ is a strategyproof rule, this implies that the analysis of \citet{BCHL+15} is tight and the harmonic rule has distortion exactly $\Theta(\sqrt{m \log m})$.

For convenience, we include an explicit proof of the lower bound that does not use strategyproofness.

\begin{theorem}
	\label{thm:LB-harmonic-rule-utilitarian}
	$\D_m(\HR,\utilityclass^{\unitsum}) = \Omega \left(\sqrt{m \log m}\right)$.
\end{theorem}

\begin{proof}
	Consider the preference profile $\profile$ with $n=m-1$ agents (the construction also works when $n$ is a multiple of $m-1$), in which each agent places a distinguished alternative $a^*$ at position $k=\sqrt{\frac{m}{2H_m}}$ (for simplicity, assume this is an integer) and the remaining alternatives are arranged cyclically in the remaining positions so that every remaining alternative appears in every remaining position once. Consider a consistent utility profile $\utilities$ in which each agent has utility $1/k$ for her $k$ most preferred alternatives and utility $0$ for all other alternatives. 
	
First, note that the optimal social welfare is $\SW(a^*,\utilities) = (m-1) \cdot \frac{1}{k} \ge \frac{m}{2k}$, where the last transition holds for $m \ge 2$. In contrast, for any $a \in A \setminus\set{a^*}$, we have $\SW(a,\utilities) = (k-1) \cdot \frac{1}{k} \le 1$ because alternative $a$ is among the top $k$ alternatives of precisely $k-1$ agents. 

Finally, the harmonic score of $a^*$ is $\harm(a^*,\profile) = (m-1) \cdot 1/k$, meaning that the harmonic rule $\HR$ selects $a^*$ with probability $\xx(a^*) = \frac{1}{2m} + \frac{1/k}{2H_m} \le \max(\frac{1}{m},\frac{1}{k H_m})$. Hence, the distortion of $\HR$ satisfies
\begin{align*}
	\D_m(\HR,\utilityclass^{\unitsum}) &\ge \frac{\SW(a^*,\utilities)}{\xx(a^*) \cdot \SW(a^*,\utilities) + (1-\xx(a^*)) \cdot 1}\\
	&\ge \frac{\SW(a^*,\utilities)}{\xx(a^*) \cdot \SW(a^*,\utilities) + 1}\\
	&= \frac{1}{\xx(a^*) + \frac{1}{\SW(a^*,\utilities)}}\\
	&\ge \frac{1}{\max\left(\frac{1}{m},\frac{1}{k H_m}\right) + \frac{2k}{m}}\\
	&\ge \frac{1}{2 \max\left(\frac{1}{m},\frac{1}{k H_m}, \frac{2k}{m} \right)} \\
	&= \min\left(\frac{m}{2}, \frac{k H_m}{2}, \frac{m}{4k} \right).
\end{align*}
Setting $k = \sqrt{\frac{m}{2H_m}}$, we get that the distortion is at least $\min(m/2,\sqrt{m H_m/8}) = \Omega(m \log m)$. 
\end{proof}

Next, we analyze the distortion of $\HR$ for unit-range and approval utilities. Strikingly, while the distortion of $\HR$ for unit-sum utilities is only a sublogarithmic factor worse than the best possible distortion, we find that its distortion for approval and unit-range utilities is $\Theta(m^{2/3} \log^{1/3} m)$, which is worse than the best possible distortion of $\Theta(\sqrt{m})$ for these utility classes by a polynomial factor. This contrast can be explained due to a result of \citet{filos2014truthfulrange} and \citet{lee2019maximization}: for unit-range utilities, the best distortion achieved by any \emph{strategyproof} voting rule (see \Cref{def:strategyproofness}) is $\Theta(m^{2/3})$. Because $\HR$ is known to be strategyproof~\citep{bhaskar2018truthful}, it must have distortion $\Omega(m^{2/3})$; further, its distortion for unit-range utilities is still only a sublogarithmic factor worse than that of the best \emph{strategyproof} voting rule. Note that our stable lottery rule $\SLR$ achieves $\Theta(\sqrt{m})$ distortion for both unit-sum and approval utilities. 

\begin{theorem}\label{thm:hr-approval}
	The distortion of the harmonic rule with respect to the class of unit-range and of approval utilities is $\D_m(\HR,\utilityclass^{\unitrange}) = \D_m(\HR,\utilityclass^{\approval}) = \Theta(m^{2/3} \log^{1/3} m)$. 
\end{theorem}
\begin{proof}
We begin by proving the upper bound.

\paragraph{Upper bound.} 
By \Cref{lem:approval-unit-range}, for proving the upper bound for unit-range utilities, it suffices to consider approval utilities.
Fix an arbitrary profile $\profile$, and let $\utilities$ be some consistent utility profile of approval utilities. For each agent $i \in N$, let $r_i = \sum_{a\in A} u_i(a) \ge 1$ be the number of alternatives approved by $i$. Let $a^* \in \argmax_{a \in A} \SW(a,\utilities)$ be an optimal alternative, and let $\xx = \HR(\profile)$ be the distribution selected by the harmonic rule.

Let $\tau$ be a threshold value to be set later. Consider two cases.
	
\begin{description}
\item[Case 1:] Suppose $\harm(a^*) \ge \tau$. Then $\xx(a^*) \ge \frac{1}{2} \cdot \frac{\tau}{n \cdot H_m}$ and so $\SW(\xx,\utilities) \ge \frac{1}{2} \cdot \frac{\tau}{n \cdot H_m} \SW(a^*,\utilities)$. Thus $\D(\xx, \utilities) = \SW(a^*,\utilities)/\SW(\xx,\utilities) \le 2nH_m/\tau$. 
\medskip
\item[Case 2:] Suppose $\harm(a^*) \le \tau$. Let $Y = \set{i \in N : u_i(a^*) = 1}$ be the set of agents approving $a^*$. Note that $\SW(a^*,\utilities) = |Y|$. Because $\xx(a) \ge 1/(2m)$ for each $a\in A$, we have $\SW(\xx,\utilities) \ge \sum_{i \in N} \frac{r_i}{2m}$. 
Thus 
\begin{equation}
\label{eq:case2-bound}
\textstyle
\D(\xx, \utilities) = \SW(a^*,\utilities)/\SW(\xx,\utilities) \le2m|Y|/(\sum_{i\in N} r_i).
\end{equation} 
We will upper bound this quantity in two different ways.

First, we have $\D(\xx, \utilities) \le 2 m |Y|/n$ because $r_i \ge 1$ for each $i \in N$. 

Second, we can observe that 
\begin{equation}
\label{eq:case2-am-hm-consequence}
\tau \ge \harm(a^*) \ge \sum_{i \in Y} \frac{1}{r_i} \ge \frac{|Y|^2}{\sum_{i \in Y} r_i} \ge \frac{|Y|^2}{\sum_{i\in N} r_i},
\end{equation}
where the first inequality is due to the assumption of Case 2, the second inequality is because every $i \in Y$ ranks $a^*$ among the first $r_i$ positions, and the third inequality is the inequality of arithmetic and harmonic means.
Rewriting \eqref{eq:case2-am-hm-consequence}, we have $|Y|/(\sum_{i\in N} r_i) \le \tau / |Y|$.
Plugging this into \eqref{eq:case2-bound}, we see that $\D(\xx, \utilities) \le 2m\tau/|Y|$. 

Combining the two bounds, and using $\min(x,y) \le \sqrt{xy}$, we see that in Case 2 we have
\[
\D(\xx, \utilities) \le \min\left(\frac{2m|Y|}{n},\frac{2m\tau}{|Y|}\right) \le 2m \, \sqrt{\tfrac{\tau}{n}}.
\]
\end{description}

\noindent
Finally, combining Case 1 and Case 2, we can see that the distortion is at most
\[
\D(\xx, \utilities) \le \max\left(2nH_m/\tau, \:2m\, \sqrt{\tfrac{\tau}{n}}\right).
\]
Setting $\tau = n \cdot (H_m/m)^{2/3}$ yields the optimal upper bound of $2 H_m^{1/3} m^{2/3}$.

\paragraph{Lower bound.}	 Assume $m \ge 2$ without loss of generality. Let $t = (H_m/m)^{1/3}$ and $r = 1/t = (m/H_m)^{1/3}$. Choose an arbitrary alternative $a^* \in A$ and construct a preference profile as follows: 
	\begin{itemize}
	\item Alternative $a^*$ is ranked $r$-th by $n \cdot t$ ``special'' agents and $m$-th by the remaining $n \cdot (1-t)$ ``ordinary'' agents. 
	\item The remaining preferences are filled arbitrarily subject to the condition that each of the remaining $m-1$ alternatives appear as the top choice of $n \cdot (1-t)/(m-1)$ ordinary agents and in the first $r-1$ positions in the preference rankings of $n \cdot t \cdot (r-1)/(m-1)$ special agents. 
	\end{itemize}

We set a consistent utility profile as follows:
\begin{itemize}
	\item Every special agent has utility $1$ for her top $r$ alternatives and $0$ for the rest.
	\item Every ordinary agent has utility $1$ for her top alternative and $0$ for the rest.
\end{itemize}

Let us analyze the harmonic scores and welfare of various alternatives. For the chosen alternative $a^*$, we have
\[
\harm(a^*) = n \cdot \left(\frac{t}{r} + \frac{1-t}{m}\right) \le n \cdot (t^2+1/m) \le 2nt^2,
\]
where the final transition uses the fact that $1/m \le t^2 = (H_m/m)^{2/3}$. Based on this, we get that the probability of $a^*$ being chosen under $\HR$ is 
\[
\Pr(a^*) \le \frac{1}{2} \cdot \frac{2t^2}{H_m} + \frac{1}{2m} \le \frac{3t^2}{2H_m},
\]
where the final transition uses the fact that $1/m \le t^2/H_m$. Next, the social welfare of $a^*$ is 
\[
\SW(a^*,\utilities) = n \cdot t,
\]
whereas the social welfare of every other alternative $a \in A \setminus \set{a^*}$ is
\[
\SW(a,\utilities) = \frac{n \cdot t \cdot (r-1)}{m-1} + \frac{n \cdot (1-t)}{m-1} \le \frac{n}{m-1} \cdot (tr+1) \le \frac{4n}{m},
\]
where the final transition uses $rt=1$ and $m-1 \ge m/2$. 

Hence, we get that the distortion of $\HR$ is at least
\begin{align*}
	&\frac{n \cdot t}{n \cdot t \cdot \frac{3t^2}{2H_m} + \frac{4n}{m} \cdot 1} = \frac{2 \cdot m \cdot t}{11} = \frac{2}{11} \cdot H_m^{1/3} \cdot m^{2/3},
\end{align*}
as needed.
\end{proof}

\subsection{Proportional Fairness}\label{sec:hr-pf}

We show that the harmonic rule is $\Theta(\sqrt{m \log m})$-proportionally fair. Since we have shown that the best voting rule is $\Theta(\log m)$-proportionally fair, the harmonic rule is worse by a polynomial factor. Later, we explain this contrast once again via strategyproofness of the harmonic rule, by proving that if an $\alpha$-proportionally fair voting rule is strategyproof, then $\alpha = \Omega(\sqrt{m})$. Hence, the harmonic rule is only a sublogarithmic factor worse than the best strategyproof voting rule according to the proportional fairness metric.

\begin{theorem}
	\label{thm:hr-pf}
	The harmonic rule is $\Theta(\sqrt{m \log m})$-proportionally fair.
\end{theorem}
\begin{proof}
Let us begin by proving the upper bound. Let $\profile$ be a preference profile and let $\xx = \HR(\profile)$ be the probability distribution returned by $\HR$ on $\profile$. From \Cref{lem:pf-approval-simplification}, recall that
\begin{equation}\label{eqn:harm-pf}
\PF(\xx,\profile) = \max_{a \in A} \frac{1}{n} \cdot \sum_{i \in N} \frac{1}{\xx(h_i(a))},
\end{equation}
where $h_i(a) = \set{b : b \succeq_i a}$ is the set of alternatives that agent $i$ ranks at least as high as $a$. Let $a^*$ denote the arg max of the right-hand side of \eqref{eqn:harm-pf}.

For $r \in [m]$, let $\alpha_r$ denote the fraction of agents who rank $a^*$ in position $r$. Note that $\sum_{r=1}^m \alpha_r = 1$. Further, the harmonic score of $a^*$ is given by $\harm(a^*) = n \cdot \sum_{r=1}^m \alpha_r/r$. We consider two cases.

\medskip\noindent\emph{Case 1.} Suppose $\sum_{r=1}^m \frac{\alpha_r}{r} \ge \sqrt{H_m/m}$. Then, $\harm(a^*) \ge n \sqrt{H_m/m}$. Hence, 
\[
\xx(h_i(a^*)) \ge \xx(a^*) \ge \frac{1}{2} \cdot \frac{1}{\sqrt{m H_m}}.
\]
Plugging this into \eqref{eqn:harm-pf}, we get that $\PF(\xx,\profile) \le 2 \sqrt{m H_m}$, as desired.

\medskip\noindent\emph{Case 2.} Suppose $\sum_{r=1}^m \frac{\alpha_r}{r} \le \sqrt{H_m/m}$. Note that $\xx(a) \ge \frac{1}{2m}$ for every alternative $a \in A$. Hence, if agent $i$ ranks $a^*$ in position $r$, we have $\xx(h_i(a)) \ge \frac{r}{2m}$. Plugging this into \eqref{eqn:harm-pf}, we get
\[
\PF(\xx,\profile) \le \sum_{r=1}^m \frac{(2m) \cdot \alpha_r}{r} \le 2 \sqrt{m H_m},
\]
as desired.

Next, we prove the lower bound. Fix a special alternative $a^*$. Construct a preference profile $\profile$ in which there are $n=m-1$ agents. Alternative $a^*$ is ranked in position $r = \sqrt{m/H_m}$ by all the agents. The other alternatives are placed in the remaining positions in a cyclic manner, so that every other alternative appears in every remaining position exactly once. Let $\xx = \HR(\profile)$ be the probability distribution returned by the harmonic rule on this profile. Note that  
\[
\xx(a^*) = \frac{1}{2} \frac{\harm(a^*)}{n H_m} + \frac{1}{2} \frac{1}{m} = \frac{1}{2} \frac{1}{r H_m} + \frac{1}{2} \frac{1}{m} \le \frac{1}{\sqrt{m H_m}}, 
\]
where the last inequality holds because $m \ge H_m$. By symmetry, the remaining probability is equally distributed among the remaining alternatives. Hence, we have $\xx(a) \le 1/(m-1)$ for all $a \in A \setminus \set{a^*}$. 

Next, fix a utility profile $\utilities$ in which every agent $i$ has utility $1$ for her $r$ most favorite alternatives. Note that for every agent $i \in N$, we have
\[
u_i(\xx) \le \frac{1}{\sqrt{m H_m}} + \frac{r-1}{m-1} \le \frac{1}{\sqrt{m H_m}} + \frac{r}{m} = \frac{2}{\sqrt{m H_m}}.
\]
In contrast, $u_i(a^*) = 1$ for all agents $i \in N$. Hence,
\[
\PF(\xx,\utilities) \ge \frac{1}{n} \sum_{i \in N} \frac{u_i(a^*)}{u_i(\xx)} \ge \frac{\sqrt{m H_m}}{2},
\]
as desired.
\end{proof}

\section{Proportional Fairness of Strategyproof Voting Rules}
\label{sec:truthful-pf}

When considering distortion with respect to utilitarian social welfare under unit-sum utilities, the harmonic rule $\HR$ provides a distortion that is only a sublogarithmic factor worse than the optimum.
Because the harmonic rule is strategyproof, in this context, strategyproofness comes at little cost.

In this section, we show that strategyproofness imposes a much larger cost with respect to proportional fairness. While we have shown that $O(\log m)$-proportional fairness is possible, we prove that every strategyproof voting rule can only be $\Omega(\sqrt{m})$-proportionally fair. Note that since the harmonic rule is $\Theta(\sqrt{m \log m})$-proportionally fair, it is at most a sublogarithmic factor worse than the best strategyproof voting rule. 

To prove our lower bound, we need to understand the class of strategyproof rules in detail. We start with a formal definition.

\begin{definition}[Strategyproofness]
	\label{def:strategyproofness}
	A voting rule $f$ is called \emph{strategyproof} (also known as \emph{truthful}) if no agent can increase her utility by misreporting her vote. Formally, for any preference profile $\profile$, any agent $i \in N$, any utility function $u_i$ consistent with $\pref_i$, and any ranking of the alternatives $\pref'_i$, we must have $u_i(f(\profile)) \ge u_i(f((\pref'_i,\profile_{-i})))$, where $(\pref'_i,\profile_{-i})$ is the preference profile obtained by replacing the vote of agent $i$ in $\profile$ by $\pref'_i$. 
\end{definition}

Before proving the result, we need to introduce several other definitions. First, we define two well-known and mild properties of (probabilistic) voting rules. 

\begin{definition}[Anonymity]
	A voting rule $f$ is called \emph{anonymous} if its outcome does not depend on the identities of the agents. Formally, for any preference profile $\profile$ and any permutation of the agents $\pi^N : N \to N$, we must have $f(\pi^N \circ \profile) = f(\profile)$, where $\pi ^N \circ \profile= (\pref_{\pi^N(i)})_{i \in N}$ is the profile obtained by permuting the votes in $\profile$ according to $\pi^N$. 
\end{definition}

\begin{definition}[Neutrality]
	A voting rule $f$ is called \emph{neutral} if its outcome does not depend on the names of the alternatives. Formally, for any preference profile $\profile$ and any permutation of the alternatives $\pi^A : A \to A$, we must have $f(\pi^A \circ \profile) = \pi^A \circ f(\profile)$, where $\pi^A \circ \profile$ is the profile obtained by permuting the alternatives in each vote $\pref_i$ according to $\pi^A$, and $\pi^A \circ f(\profile)$ is the distribution obtained by permuting the names of the alternatives in $f(\profile)$ according to $\pi^A$. 
\end{definition}

Next, we introduce two classes of voting rules, following the work of \citet{Bar78}. 

\begin{definition}[Point-Voting Schemes]
	A voting rule $f$ is called a \emph{point-voting scheme} if there exists a vector $\ww = (w_1,\ldots,w_m)$ with $w_1 \ge w_2 \ge \cdots \ge w_m \ge 0$ and $\sum_{r \in [m]} w_r = 1$, such that for every preference profile $\profile$, writing $\xx = f(\profile)$, we have 
	\[
	\textstyle
	\xx(a) = \frac1n \sum_{i \in N} w_{\pref_i(a)} \quad\text{for all $a \in A$.}
	\]
\end{definition}

Informally, a point-voting scheme, parametrized by the vector $\ww$, resembles the positional scoring rule parametrized by the same score vector $\vec{w}$, except that the positional scoring rule would choose the alternative $a$ with the highest total score $\sum_{i \in N} w_{\pref_i(a)}$ whereas the point-voting scheme chooses each alternative $a$ with probability proportional to its score. Another way to view a point-voting scheme is that it chooses an agent uniformly at random and then chooses her $r$-th ranked alternative with probability $w_r$, for each $r \in [m]$. 

Note that the first half of the harmonic rule $\HR$, which chooses each alternative with probability proportional to its harmonic score, is a point-voting scheme. In fact, the entire harmonic rule $\HR$ is a point-voting scheme, obtained using $w_r = (1/r)+(H_m/m)$ for each $r \in [m]$; this can alternatively be confirmed by noting that the set of point-voting schemes is closed under convex combinations and the second half of the harmonic rule, which picks an alternative uniformly at random, is obviously a point-voting scheme with $w_r = 1/m$ for each $r \in [m]$.

\begin{definition}[Supporting-Size Schemes]
	A voting rule $f$ is called a \emph{supporting-size scheme} if there exists a vector $\vzz = (z_0,\ldots,z_n)$ with 
	\begin{itemize}
		\item $z_n \ge z_{n-1} \ge \cdots \ge z_0$, and
		\item $z_k + z_{n-k} = 1$ for each $k \in [n] \cup \set{0}$
	\end{itemize}
	such that for every preference profile $\profile$, writing $\xx = f(\profile)$, we have 
	\[
	\xx(a) = \frac{1}{\binom{m}{2}} \sum_{b \in A \setminus \set{a}} z_{V(a, b)} \quad\text{for all $a \in A$, where $V(a, b) = |\set{i \in N : a \succ_i b}|$.} 
	\]
\end{definition}

In other words, $f$ chooses a pair of alternatives $(a,b)$ uniformly at random. Writing $k$ for the number of voters that prefer $a$ over $b$, the rule $f$ then chooses $a$ with probability $z_k$ and chooses $b$ with probability $z_{n-k}$.

\citet{Bar78} proved the following characterization result.

\begin{proposition}\label{prop:char-anon-neut-sp}
	A voting rule is anonymous, neutral, and strategyproof if and only if it is a probability mixture of a point-voting scheme and a supporting-size scheme.
\end{proposition}

The reason this result is useful is that by analyzing the best objective value (distortion or proportional fairness) achievable by any mixture of a point-voting scheme and a supporting-size scheme, we also obtain the best objective value achievable by any anonymous, neutral, and strategyproof voting rule. Could a strategyproof voting rule that violates anonymity and/or neutrality achieve a better objective value? For distortion, \citet{filos2014truthfulrange} prove that this is not the case, and this observation was used by them and by \citet{bhaskar2018truthful} to derive the aforementioned lower bounds on the distortion of any strategyproof voting rule with respect to the unit-range and unit-sum utility classes, respectively. It is easy to see that the same observation holds for proportional fairness as well. 

\begin{lemma}\label{lem:pf-anon-neut}
For every strategyproof voting rule $f$, there exists an anonymous, neutral, and strategyproof voting rule $f'$ such that $\D^{\PF}_m(f') \le \D^{\PF}_m(f)$. 
\end{lemma}
\begin{proof}
Like \citet{filos2014truthfulrange}, we consider a strategyproof voting rule $f$ and construct a voting rule $f'$ which works as follows: given an input preference profile $\profile$, it applies a uniformly random permutation of the agents $\pi^N$ and an independently chosen uniformly random permutation of alternatives $\pi^A$ to $\profile$, then applies rule $f$ on the resulting profile $\pi^A \circ \pi^N \circ \profile$, and finally applies the inverse of $\pi^A$, denoted $(\pi^A)^{-1}$, on the resulting distribution to revert the change of names of alternatives. 
	
\citet{filos2014truthfulrange} argue that if $f$ is strategyproof, then $f'$ is anonymous, neutral, and strategyproof. Further, $\D_m(f',\utilityclass^{\unitrange}) \le \D_m(f,\utilityclass^{\unitrange})$ (and $\D_m(f',\utilityclass^{\unitsum}) \le \D_m(f,\utilityclass^{\unitsum})$ using the same argument). We want to show that $\D^{\PF}_m(f') \le \D^{\PF}_m(f)$ as well. The argument for this is slightly more involved because, unlike the social welfare function, proportional fairness is non-linear. Crucially, we use the fact that $\D^{\PF}(\xx,\profile)$ is convex in $\xx$, as we observed after the proof of \Cref{lem:pf-approval-simplification}.
	
Take any preference profile $\profile$, and let $\xx = f'(\profile)$. Let $\Pi^N$ and $\Pi^A$ denote the set of permutations of agents and alternatives, respectively. Note that
\[
\xx = \frac{1}{|\Pi^N| \cdot |\Pi^A|} \textstyle\sum_{\pi^N \in \Pi^N, \pi^A \in \Pi^A} (\pi^A)^{-1} \circ f(\pi^A \circ \pi^N \circ \profile).
\]

Hence, we have 
\begin{align*}
\D^{\PF}(\xx,\profile) &\le \frac{1}{|\Pi^N| \cdot |\Pi^A|} \textstyle\sum_{\pi^N \in \Pi^N, \pi^A \in \Pi^A} \D^{\PF}((\pi^A)^{-1} \circ f(\pi^A \circ \pi^N \circ \profile),\profile)\\
&\le \textstyle\max_{\pi^N \in \Pi^N, \pi^A \in \Pi^A} \D^{\PF}((\pi^A)^{-1} \circ f(\pi^A \circ \pi^N \circ \profile),\profile)\\
&= \textstyle\max_{\pi^N \in \Pi^N, \pi^A \in \Pi^A} \D^{\PF}((\pi^A)^{-1} \circ f(\pi^A \circ \pi^N \circ \profile),\pi^N \circ \profile)\\
&= \textstyle\max_{\pi^N \in \Pi^N, \pi^A \in \Pi^A} \D^{\PF}(f(\pi^A \circ \pi^N \circ \profile),\pi^A \circ \pi^N \circ \profile)\\
&\le \max_{\profile'} \D^{\PF}(f(\profile'),\profile') = \D^{\PF}_m(f),
\end{align*}
where the first transition is due to convexity of the proportional fairness objective, the second transition upper-bounds an average by the maximum, the third transition uses the fact that proportional fairness is an anonymous objective (i.e., permuting the votes does not change the proportional fairness value of a distribution on the preference profile), the fourth transition uses the fact that proportional fairness is a neutral objective (i.e., permuting the names of alternatives in both the distribution and the preference profile keeps the proportional fairness value unchanged), and the final transition upper-bounds the maximum over a subset of preference profiles by a maximum over all preference profiles. Since this holds for each preference profile $\profile$, we have $\D^{\PF}_m(f') \le \D^{\PF}_m(f)$, as required. 
\end{proof}

We are now equipped to prove a lower bound on proportional fairness of any strategyproof voting rule.
\begin{theorem}
	\label{thm:strategyproof-lb-pf}
	For every strategyproof voting rule $f$, we have $\D^{\PF}_m(f) = \Omega(\sqrt{m})$. 
\end{theorem}
\begin{proof}
Let $f$ be a strategyproof voting rule.
Due to \Cref{lem:pf-anon-neut}, we may assume that $f$ is anonymous and neutral. Due to \Cref{prop:char-anon-neut-sp}, $f$ is a probability mixture that implements some point-voting scheme characterized by $\ww$ with probability $p \in [0,1]$ and some supporting-size scheme characterized by $\vzz$ with probability $1-p$.

Let $\alpha = \D^{\PF}_m(f)$. We will show that $\alpha = \Omega(\sqrt{m})$. Similarly to \citet{bhaskar2018truthful}, we construct a sequence of profiles $\profile^r$, one for each $r \in [m]$, and show that $f$ must select a distribution that is at least $\Omega(\sqrt{m})$-proportionally fair on at least one of these profiles, so that $\alpha = \Omega(\sqrt{m})$. 
	
Fix any $r \in [m]$ and construct the preference profile $\profile^r$ with $n=m-1$ votes,\footnote{The profile can easily be replicated to make $n$ any multiple of $m-1$.} such that a special alternative $a^*$ appears in position $r$ in all the votes, while the remaining $m-1$ alternatives fill the remaining positions in a cyclic order. Let $\xx^r = f(\profile^r)$. Then, 
\begin{align}
\xx^r(a^*) &= p \cdot \frac{1}{n} \cdot \sum_{i \in N} w_{\pref_i(a^*)} + (1-p) \cdot \frac{1}{\binom{m}{2}} \cdot \sum_{a \in A\setminus\set{a^*}} z_{V(a^*, a)} \nonumber \\
&\le 1 \cdot \frac{1}{n} \cdot \sum_{i \in N} w_r + 1 \cdot \frac{1}{\binom{m}{2}} \cdot \sum_{a \in A\setminus\set{a^*}} 1 	\nonumber \\
&= w_r + \frac{2}{m}. \label{eqn:pf-prob-ub}
\end{align}
	
As argued by \citet{bhaskar2018truthful}, this observation would effectively allow us to ignore the impact of the supporting-size scheme, which can only select the ``desired'' alternative $a^*$ with a small probability of $2/m$, and focus on the impact of the point-voting scheme. Let us lower-bound the proportional fairness of $\xx^r$ on $\profile^r$. We have
\begin{align*}
\D^{\PF}(\xx^r,\profile^r) &= \max_{a \in A} \frac{1}{n} \sum_{i \in  N} \frac{1}{\xx^r(h_i(a))} \tag{by \Cref{lem:pf-approval-simplification}}\\
&\ge \frac{1}{n} \sum_{i \in  N} \frac{1}{\xx^r(h_i(a^*))}\\
&\ge \frac{n}{\sum_{i \in  N} \xx^r(h_i(a^*))} \tag{by the AM-HM inequality}\\
&= \frac{n}{n \cdot \left(\xx^r(a^*) + \frac{r-1}{m-1} \cdot (1-\xx^r(a^*))\right)}\\
&\ge \frac{1}{\xx^r(a^*) +  \frac{r}{m} \cdot (1-\xx^r(a^*))}\\
&\ge \frac{1}{\xx^r(a^*) +  \frac{r}{m}},
\end{align*} 
where the fourth transition holds because anonymity and neutrality of $f$ implies $\xx^r(a) = \frac{1-\xx^r(a^*)}{m-1}$ for each $a \in A\setminus\set{a^*}$. By definition of $\alpha$, we have $\alpha \ge \D^{\PF}(\xx^r,\profile^r)$. Thus it follows that
\[
\xx^r(a^*) + \frac{r}{m} \ge \frac{1}{\alpha}.
\]

Using \eqref{eqn:pf-prob-ub}, we have
\[
w_r \ge \max\left(\frac{1}{\alpha} - \frac{r+2}{m},0\right).
\]

Summing over all $r \in [m]$, we get
\begin{align*}
	1 = \sum_{r \in [m]} w_r &\ge \sum_{r \in [m]} \max\left(\frac{1}{\alpha} - \frac{r+2}{m},0\right)\\
	&= \sum_{r=1}^{\lfloor\frac{m}{\alpha}\rfloor-2} \left(\frac{1}{\alpha}-\frac{r+2}{m}\right)\\
	&\ge \frac{\frac{m}{\alpha}-3}{\alpha} - \frac{\frac{m}{\alpha}\left(\frac{m}{\alpha}+1\right)- {6}}{2m}\\
	&\ge \frac{m}{2\alpha^2} - \frac{7}{2\alpha} + \frac{6}{2m},
\end{align*}
where the fourth transition uses $m/\alpha-1 \le \lfloor m/\alpha \rfloor \le m/\alpha$. The above expression simplifies to
\[
(2m-6) \alpha^2 + 7m\alpha - m^2 \ge 0,
\]
which shows
\[
\alpha \ge \frac{-7m + \sqrt{49m^2 + 4m^2(2m-6)}}{2(2m-6)} = \Omega(\sqrt{m}),
\]
as needed. 
\end{proof}

\section{Proportional Fairness via Approximately Stable Committees}\label{app:scr}

As indicated in \Cref{sec:fairness}, we prove that a rule similar to our stable lottery rule ($\SLR$), which uses a deterministic committee satisfying approximate stability instead of a lottery over committees satisfying exact stability, is $O(\sqrt{m})$-proportionally fair, and therefore achieves $O(\sqrt{m})$ distortion with respect to the Nash welfare. Let us first formally introduce approximate stability for committees.  

\begin{definition}[Approximately Stable Committees]
For a committee $X$ with $|X| = k$ and an alternative $a^*$, recall that $V(a^*,X) = |\{ i \in N : a^* \succ_i X\}|$ denotes the number of agents who prefer $a^*$ to all alternatives in $X$. We say that $X$ is $c$-\emph{stable} if for all alternatives $a^* \not\in X$, we have $V(a^*,X) < c \cdot \frac{n}{k}$. \end{definition}

Note that $1$-stable committees are precisely the stable committees introduced in \Cref{sec:stable}. As mentioned in that section, there exist preference profiles and sizes $k$ where no stable committee of size $k$ exists~\citep[Thm.~4]{jiang2020approximately}. However, by derandomizing the stable lottery of \citet{cheng2020group}, \citet{jiang2020approximately} proved the following:

\begin{theorem}[\citealp{jiang2020approximately}]
	Given any ranked preference profile and $k \in [m]$, a $16$-stable committee of size $k$ exists and a $(16+\epsilon)$-stable committee of size $k$ can be computed in $\poly(n,m,1/\epsilon)$ time for sufficiently small constant $\epsilon > 0$. 
\end{theorem}

Let us introduce a voting rule that uses an approximately stable committee in the same manner in which the rule $\SLR$ from \Cref{sec:distortion} uses an exactly stable lottery. Note that despite the use of a deterministic committee, the rule is still probabilistic in the end. 

\begin{definition}[$c$-Stable Committee Rule, $\cSCR$]
Let $X$ be a $c$-stable committee of size $k = \sqrt{m}$. The $c$-Stable Lottery Rule ($\cSCR$) works as follows: With probability $1/2$, choose an alternative uniformly at random from $X$, and with probability $1/2$, choose an alternative uniformly at random from $A$. Therefore, each alternative $a \in A$ is selected with probability $\xx(a) = \frac{1}{2\sqrt{m}} \cdot \id[a \in X] + \frac{1}{2m}$, where $\id$ is the indicator function.
\end{definition}

Next, we prove that $\cSCR$ is $O(\sqrt{m})$-proportionally fair when $c$ is constant.

\begin{theorem}\label{thm:pf-scr}
	We have $\D^{\PF}_m(\cSCR) = O(c \cdot \sqrt{m})$. 
\end{theorem}
\begin{proof}
For constant $c$, consider the $\cSCR$ rule. Fix an arbitrary preference profile $\profile$. Let $X$ be the $c$-stable committee of size $\sqrt{m}$ that $\cSCR$ uses to output distribution $\xx$ on this profile. We want to prove that $\D^{\PF}(\xx,\profile) = O(\sqrt{m})$. 
	
By \Cref{lem:pf-approval-simplification}, we have that
\begin{equation}\label{eqn:pf-scr}
\D^{\PF}(\xx,\profile) = \max_{a \in A} \frac{1}{n} \sum_{i \in N} \frac{1}{\xx(h_i(a))} = \frac{1}{n} \sum_{i \in N} \frac{1}{\xx(h_i(a^*))},
\end{equation}
where $\xx(h_i(a))$ is the probability placed on the set of alternatives $h_i(a) = \set{a' : a' \succcurlyeq_i a}$ under $\xx$ and $a^*$ is taken to be an arg max. 

Let $S \subseteq N$ denote the set of $V(a^*,X)$ many agents who prefer $a^*$ to every alternative in $X$. Because $X$ is $c$-stable, we know that $|S| = V(a^*,X) \le c \cdot n/\sqrt{m}$. By definition of $S$, each agent $i \in N \setminus S$ satisfies $h_i(a^*) \cap X \neq \emptyset$, implying that $\xx(h_i(a^*)) \ge \frac{1}{2\sqrt{m}}$. For each agent $i \in S$, we have $\xx(h_i(a^*)) \ge \xx(a^*) \ge \frac{1}{2m}$. Plugging these lower bounds into \eqref{eqn:pf-scr}, we get
\begin{align*}
	\D^{\PF}(\xx,\profile) &= \frac{1}{n} \sum_{i \in N} \frac{1}{\xx(h_i(a^*))}\\
	&= \frac{1}{n} \left( \sum_{i \in S} \frac{1}{\xx(h_i(a^*))} + \sum_{i \in N \setminus S} \frac{1}{\xx(h_i(a^*))}\right)\\
	&\le \frac{1}{n} \left( |S| \cdot 2m + |N\setminus S| \cdot 2\sqrt{m} \right)\\
	&\le \frac{1}{n} \left( c \cdot \frac{n}{\sqrt{m}} \cdot 2m + n \cdot 2\sqrt{m} \right)\\
	&= 2 \cdot (c+1) \cdot \sqrt{m} = O(c \cdot \sqrt{m}). 
\end{align*}
This completes the proof.
\end{proof}

Note that an upper bound on proportional fairness also applies to distortion with respect to the Nash welfare. 
\begin{corollary}
	The distortion of $\cSCR$ with respect to the Nash welfare satisfies \\ $\D^{\NW}_m(\cSCR,\utilityclass^{\all}) = O(c \cdot \sqrt{m})$. 
\end{corollary}

For distortion with respect to the utilitarian social welfare, the proof of \Cref{thm:distortion_ub} can easily be modified to show that for each constant $c$, the distortion of  $\cSCR$ is $O(\sqrt{m})$ for balanced utilities (and therefore also for unit-sum and unit-range utilities). 
\begin{theorem}
	\label{thm:distortion_ub_stable_committee}
	On the utility class $\utilityclass^{\balanced}$, the distortion of $\cSCR$ with respect to utilitarian social welfare satisfies
	$\D^{\SW}_m(\cSCR,\utilityclass^{\balanced}) = O(c \cdot \sqrt{m})$.
\end{theorem}
In the main body, we focused on the Stable Lottery Rule instead of the Stable Committee Rule because for the former it is easier to prove existence, its output is easier to compute, and due to the existence of an \emph{exact} stable lottery (as opposed to an approximately stable committee), we get a distortion upper bound of $2\sqrt{m}$, which is close to the lower bound of $\sqrt{m}/2$.

We end this section by recalling that in \Cref{sec:fairness}, we show that a better performance on proportional fairness can be achieved, using a different technique based on the minimax theorem. 

\section{The Case of Two Alternatives}\label{app:nash-two}
In this section, we analyze the case of exactly two alternatives, say, $a_1$ and $a_2$, which is an important special case, capturing referenda and pairwise comparisons. Consider a situation in which 60\% of the voters prefer $a_1$ and 40\% of the voters prefer $a_2$. Given such a split, how much probability should be placed on $a_1$ and $a_2$? The straightforward answers would be to place 100\% on $a_1$ (because this is the outcome of majority voting), or perhaps to place $60\%$ on $a_1$ (for example, because this is the outcome of random dictatorship). This section shows that the best answer need not be either of those two, but rather something more complicated. 

\begin{figure}[t]
	\centering
		\includegraphics[width=0.75\textwidth]{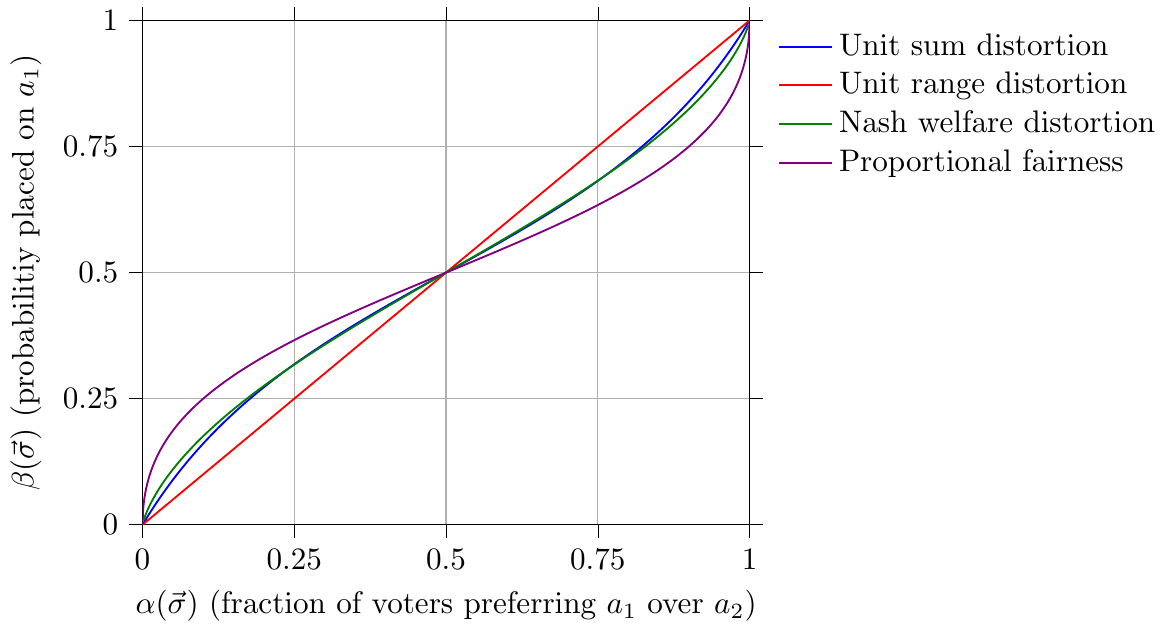}
     	\caption{The instance-optimal voting rules for the four types of distortion we consider. The plot shows the probability $\beta(\profile)$ placed on alternative $a_1$ as a function of the fraction $\alpha(\profile)$ of voters who prefer $a_1$ over $a_2$}
	\label{fig:m2-rules}
\end{figure}

For each of the objectives that we studied in this paper (distortion with unit-sum utilities, distortion with unit-range utilities, distortion with respect to Nash welfare, proportional fairness), we will explicitly write down a voting rule that, for every preference profile, selects the instance-optimal distribution. 
The result is depicted in \Cref{fig:m2-rules}. There we see that the instance-optimal rule for distortion with respect to unit range utilities is indeed random dictatorship (where the probability of an alternative is proportional to the number of voters preferring it). However, for the other three objectives, we see that output distributions that are closer to the uniform distribution perform better. This effect is strongest for proportional fairness, which therefore needs to be more ``cautious'' than random dictatorship.

We will also compute the worst-case distortion/proportional fairness of these rules and hence the best possible values obtainable in the two-alternative case. These values turn out to be $3/2$, $4/3$, $\sqrt{2}$, and $3/2$, respectively.

As there are only two possible rankings over two alternatives, we can summarize a preference profile $\profile$ by a real number $\alpha(\profile) \in [0, 1]$, which denotes the fraction of agents who prefer $a_1$ to $a_2$; then, the remaining $1 - \alpha{(\profile)}$ fraction of agents prefer $a_2$ to $a_1$. Similarly, the outcome of a voting rule on a preference profile $\profile$ can also be viewed as a real number $\beta{(\profile)} \in [0, 1]$ which is the probability placed on $a_1$. We will adapt our notation accordingly throughout this section.

We can disregard the extreme cases where $\alpha(\profile) \in \{0, 1\}$ because in such cases, one alternative is preferred by all agents ($a_1$ if $\alpha(\profile)=1$ and $a_2$ if $\alpha(\profile)=0$); the voting rule should clearly choose that alternative as it achieves the optimal distortion of $1$ with respect to all our welfare functions and utility classes as well as the optimal $1$-proportional fairness.

\subsection{Distortion With Unit-Sum Utilities}

Given a preference profile $\profile$, let $\UWTwoUnitsum$ be the  voting rule that selects $\beta(\profile) = \frac{(2 - \alpha(\profile)) \cdot \alpha(\profile)}{1 + 2\alpha(\profile) \cdot (1 - \alpha(\profile))}$. We show that this rule always returns the instance optimal distribution minimizing $\D^{\SW}(\profile,\utilityclass^{\unitsum})$.

\begin{theorem}\label{thm:two-unit-sum}
	For $m=2$ alternatives, and for any preference profile $\profile$, the voting rule $\UWTwoUnitsum$ selects a distribution $\xx$ minimizing $\D^{\SW}(\xx,\profile,\utilityclass^{\unitsum})$. 
	
	The rule achieves distortion $\D_2^{\SW}(\UWTwoUnitsum,\utilityclass^{\unitsum}) = 3/2$, which is the best possible among all voting rules. 
\end{theorem}

\begin{proof}
	Fix a preference profile $\profile$. 
	Let us write $\alpha = \alpha(\profile)$ and let
	\[
	\D^{\SW}(\alpha) = \min_{\xx \in \Delta(A)}  \D^{\SW}(\xx, \profile, \utilityclass^{\unitsum})
	\] be the best achievable distortion on profile $\profile$ with unit-sum utilities. 
	Then,
	\begin{align}
	\D^{\SW}(\alpha) &=
	\min_{\beta \in [0, 1]}
	\, \max_{\utilities \in (\utilityclass^{\unitsum})^n : \utilities \induces \profile}
	\, \frac{\max \{ \SW(a_1, \utilities), \SW(a_2, \utilities) \}}{\beta \cdot \SW(a_1, \utilities) + (1 - \beta) \cdot \SW(a_2, \utilities)} \nonumber\\
	&= \min_{\beta \in [0,1]} \max\bigg\{\max_{\utilities \in (\utilityclass^{\unitsum})^n : \utilities \induces \profile} \, \frac{\SW(a_1,\utilities)}{\beta \cdot \SW(a_1, \utilities) + (1 - \beta) \cdot \SW(a_2, \utilities)} ,\nonumber\\
	&\qquad\qquad\qquad \max_{\utilities \in (\utilityclass^{\unitsum})^n : \utilities \induces \profile} \, \frac{\SW(a_2,\utilities)}{\beta \cdot \SW(a_1, \utilities) + (1 - \beta) \cdot \SW(a_2, \utilities)}\bigg\}.\label{eqn:separate}
	\end{align}
	
	\paragraph{Worst-case utilities.} In \eqref{eqn:separate}, let us analyze the worst-case utility profile $\utilities$ for the term with the numerator $\SW(a_1,\utilities)$. Using the fact that $\SW(a_1,\utilities) + \SW(a_2,\utilities) = n$ (by the unit-sum assumption), the expression of interest in \eqref{eqn:separate} is
	\[
	\frac{\SW(a_1, \utilities)}{\beta \cdot \SW(a_1, \utilities) + (1 - \beta) (n -  \SW(a_1, \utilities))}
	=
	\frac{1}{\beta + (1 - \beta)  (n - \SW(a_1, \utilities)) /  \SW(a_1, \utilities)},
	\]
	which is maximized when $\SW(a_1, \utilities)$ is maximized. To obtain the highest utilitarian welfare for $a_1$,  the $\alpha$ fraction of agents $i$ who rank $a_1$ first must have $(u_i(a_1), u_i(a_2)) = (1, 0)$ and the remaining $1- \alpha$ fraction of agents $i$ who rank $a_1$ second must have $(u_i(a_2), u_i(a_1)) = (\nicefrac{1}{2}, \nicefrac{1}{2})$. Similarly, the worst-case utility profile $\utilities$ for the term with the numerator $\SW(a_2,\utilities)$ is achieved when the $\alpha$ fraction of agents $i$ who rank $a_2$ second have $(u_i(a_1), u_i(a_2)) = (\nicefrac{1}{2}, \nicefrac{1}{2})$ and the remaining $1- \alpha$ fraction of agents $i$ who rank $a_2$ first have $(u_i(a_2), u_i(a_1)) = (1, 0)$.
	Therefore, \eqref{eqn:separate} becomes
	\begin{align*}
	\D^{\SW}(\alpha)
	&=
	\min_{\beta \in [0, 1]}
	\, \max
		\left\{
			\frac{\alpha + \frac{1- \alpha}{2}}
					{(\alpha + \frac{1-\alpha }{2})\beta + \frac{1-\alpha }{2} \cdot(1-\beta)}
					,
			\frac{\frac{\alpha}{2} + 1 - \alpha}
					{(\frac{\alpha}{2} + 1 - \alpha) (1 - \beta) + \frac{\alpha}{2} \cdot \beta}
		\right\}
	\\
	&=
	\min_{\beta \in [0, 1]}
	\, \max
		\left\{
			\frac{1}
					{\beta + \frac{(1-\alpha)/2}{\alpha + (1-\alpha )/2} \cdot(1-\beta)}
			,
			\frac{1}
					{\frac{\alpha/2}{\alpha/2 + 1 - \alpha} \cdot \beta +  1 - \beta }
		\right\}.
	\end{align*}
	The left term in the maximum is decreasing in $\beta$ while the right term is increasing. Hence, the maximum of the two is minimized when the two terms become equal, i.e.,
	\begin{align*}
	\beta + \frac{(1-\alpha)/2}{\alpha + (1-\alpha )/2} \cdot(1-\beta) &= \frac{\alpha/2}{\alpha/2 + 1 - \alpha} \cdot \beta +  1 - \beta 
	\\
	\iff
	\beta \cdot \bigg(\frac{1}{\frac{\alpha / 2}{1 - \alpha} + 1}\bigg)
	&=
	(1 - \beta) \cdot \bigg(\frac{1}{\frac{1 - \alpha}{2 \alpha } + 1}\bigg)
	\iff
	\frac{1 - \beta}{\beta} = \frac{(1-\alpha)(1 + \alpha)}{(2 - \alpha) \alpha}.
	\end{align*}
	The solution to the equation above is precisely the $\beta$ selected by $\UWTwoUnitsum$, proving that it returns the distribution with the best possible distortion on all profiles.	
	Further, at this value of $\beta$, the distortion achieved is $\D^{\SW}_2(\UWTwoUnitsum) = \max_{\alpha \in [0, 1]} \D^{\SW}(\alpha) = \nicefrac{3}{2}$, which is attained at $\alpha = \nicefrac{1}{2}$ (for which the optimal $\beta$ is also $\beta = \nicefrac{1}{2}$).
\end{proof}

From \Cref{sec:distortion}, we know that, with $m$ alternatives, the optimal distortion with respect to utilitarian welfare and unit-sum utilities is $\Theta(\sqrt{m})$, where the constant hidden in the asymptotic notation lies in $[\frac{1}{2},2]$. Interestingly, \Cref{thm:two-unit-sum} shows that the optimal distortion for $m=2$ is $\frac{3}{2} = \frac{3}{2\sqrt{2}} \cdot \sqrt{m}$, leading one to wonder whether the constant $\frac{3}{2\sqrt{2}} \approx 1.06066$ is (close to) the true constant for this problem.  

\subsection{Distortion With Unit-Range Utilities}

Given a preference profile $\profile$, let $\UWTwoUnitrange$ be the voting rule that selects $\beta(\profile) = \alpha(\profile)$. 

\begin{theorem}\label{thm:two-unit-range}
	For $m=2$ alternatives, and for any preference profile $\profile$, the voting rule $\UWTwoUnitrange$ selects a distribution $\xx$ minimizing $\D^{\SW}(\xx,\profile,\utilityclass^{\unitrange})$. 
	
	The rule achieves distortion $\D_2^{\SW}(\UWTwoUnitrange,\utilityclass^{\unitrange}) = 4/3$, which is the best possible among all voting rules. 
\end{theorem}

\begin{proof}
	Fix a preference profile $\profile$. 
	Let us write $\alpha = \alpha(\profile)$, and let
	\[
	\D^{\SW}(\alpha) = \min_{\xx \in \Delta(A)} \D^{\SW}(\xx, \profile, \utilityclass^{\unitrange})
	\] be the best achievable distortion on profile $\profile$ with unit-range utilities. 
	Then, as in the proof of \Cref{thm:two-unit-sum}, we have
	\begin{align}
	\D^{\SW}(\alpha) &=
	\min_{\beta \in [0, 1]}
	\, \max_{\utilities \in (\utilityclass^{\unitrange})^n : \utilities \induces \profile}
	\, \frac{\max \{ \SW(a_1, \utilities), \SW(a_2, \utilities) \}}{\beta \cdot \SW(a_1, \utilities) + (1 - \beta) \cdot \SW(a_2, \utilities)} \nonumber\\
	&= \min_{\beta \in [0,1]} \max\bigg\{\max_{\utilities \in (\utilityclass^{\unitrange})^n : \utilities \induces \profile} \, \frac{\SW(a_1,\utilities)}{\beta \cdot \SW(a_1, \utilities) + (1 - \beta) \cdot \SW(a_2, \utilities)} ,\nonumber\\
	&\qquad\qquad\qquad \max_{\utilities \in (\utilityclass^{\unitrange})^n : \utilities \induces \profile} \, \frac{\SW(a_2,\utilities)}{\beta \cdot \SW(a_1, \utilities) + (1 - \beta) \cdot \SW(a_2, \utilities)}\bigg\}.\label{eqn:separate-unitrange}
	\end{align}
	
	\paragraph{Worst-case utilities.} The only difference compared to the proof of \Cref{thm:two-unit-sum} is the analysis of the worst-case utility profiles in the two expressions inside the maximum in \eqref{eqn:separate-unitrange}. By \Cref{lem:approval-unit-range}, we know that the worst-case utility profile is an approval utility profile. All agents approve their first-ranked alternative and the only question is whether they also approve their second-ranked alternative. 
	
	Consider the first term inside the maximum in \eqref{eqn:separate-unitrange} with $\SW(a_1,\utilities)$ in the numerator. We want to find the utility profile $\utilities$ that maximizes this term. For the $\alpha$ fraction of agents who rank $a_1$ above $a_2$, we can assume w.l.o.g. that $u_i(a_2) = 0$ as it can only increase this term. For the rest of the agents $i$ that rank $a_2$ above $a_1$, we know $u_i(a_2) = 1$. Hence, $\SW(a_2,\utilities) = (1 - \alpha) \cdot n$. Given this fixed value of $\SW(a_2,\utilities)$, note that 
	\[
	\frac{\SW(a_1,\utilities)}{\beta \cdot \SW(a_1,\utilities) + (1-\beta) \cdot \SW(a_2,\utilities)} = \frac{1}{\beta + (1-\beta) \cdot \frac{\SW(a_2,\utilities)}{\SW(a_1,\utilities)}}
	\]
	is increasing in $\SW(a_1,\utilities)$. Hence, the term is maximized when $\SW(a_1,\utilities)$ is the highest, meaning that the agents who rank $a_2$ above $a_1$ also approve $a_1$. 
		
Similarly, for the term in \eqref{eqn:separate-unitrange} with $\SW(a_2,\utilities)$ in the numerator, the worst-case utility profile is as follows: $\alpha$ fraction of agents $i$ who rank $a_1$ first have $(u_i(a_1), u_i(a_2)) = (1, 1)$ and the remaining $1- \alpha$ fraction of agents $i$ who rank $a_2$ first have $(u_i(a_2), u_i(a_1)) = (1, 0)$. Therefore, \eqref{eqn:separate-unitrange} becomes
	\begin{align*}
		\D^{\SW}(\alpha)
		&=
		\min_{\beta \in [0, 1]}
		\, \max
		\left\{
		\frac{1}{\beta + (1 - \alpha) (1- \beta)}
		,
		\frac{1}{\alpha \beta + (1- \beta)}.
		\right\}
	\end{align*}
	The left term in the maximum is decreasing in $\beta$ while the right term is increasing. Hence, the maximum of the two terms is minimized when the two terms are equal, which yields  $\beta = \alpha$. This proves that $\UWTwoUnitrange$ returns the distribution with the best possible distortion on all profiles.	
	Further, at this value of $\beta$, the distortion achieved is $\D^{\SW}_2(\UWTwoUnitrange) = \max_{\alpha \in [0, 1]} \D^{\SW}(\alpha) = \nicefrac{4}{3}$, which is attained at $\alpha = \beta = \nicefrac{1}{2}$.
\end{proof}

\subsection{Distortion With Respect to Nash Welfare}

Given a preference profile $\profile$, let $\NashTwo$ be the voting rule that selects $\beta(\profile) = g^{-1}(\alpha(\profile))$ where $g(x) = \frac{\ln(1 - x)}{\ln(x) + \ln(1 - x)}$. This is well-defined since since $g: [0, 1] \mapsto [0, 1]$ is strictly increasing and invertible.

\begin{theorem}\label{thm:two-nash}
	For $m=2$ alternatives, and for any preference profile $\profile$, the voting rule $\NashTwo$ selects a distribution $\xx$ minimizing $\D^{\NW}(\xx,\profile,\utilityclass^{\all})$, the distortion with respect to Nash welfare. 
	
	The rule achieves distortion $\D_2^{\NW}(\NashTwo,\utilityclass^{\all}) = \sqrt{2}$ with respect to Nash welfare, which is the best possible among all voting rules. 
\end{theorem}
\begin{proof}
	Fix a preference profile $\profile$. Write $\alpha = \alpha(\profile)$, and let $\D^{\NW}(\alpha) = \min_{\xx \in \Delta(A)} \D^{\NW}(\xx, \profile, \utilityclass^{\all})$ be the best achievable distortion on the profile $\profile$. Note that 
	\begin{equation}\label{eqn:nashtwo}
		\D^{\NW}(\alpha) = \min_{\xx \in \Delta(A)} \max_{\yy \in \Delta(A)} \sup_{\utilities \in (\utilityclass^{\all})^n : \utilities \induces \profile} \left(\prod_{i \in N} \frac{u_i(\yy)}{u_i(\xx)}\right)^{1/n}.
	\end{equation}
	
	\paragraph{Reduction to approval utilities.} From \Cref{lem:nash-approval-unit-range}, we know that the worst case for distortion with respect to Nash welfare is achieved at an approval utility profile.
	Under an approval utility profile $\utilities$, the 
$\alpha$ fraction of agents $i$ who prefer $a_1$ to $a_2$ have $(u_i(a_1),u_i(a_2))$ equal to $(1, 0)$ or $(1, 1)$, and the remaining $1- \alpha$ fraction of agents $i$ have $(u_i(a_1),u_i(a_2))$ equal to $(0, 1)$ or $(1, 1)$. 
	If an agent approves both alternatives, then $\frac{u_i(\yy)}{u_i(\xx)} = 1$ regardless of $\xx$ and $\yy$. Based on these observations, we can rewrite $\D^{\NW}(\alpha)$ from \eqref{eqn:nashtwo} as
	\[
	\D^{\NW}(\alpha) = \min_{\beta \in [0, 1]} \max_{z \in [0, 1]} \left(
	\max \left\{ \frac{z}{\beta},  1  \right\}^\alpha \cdot \max \left\{ \frac{1 - z}{1 - \beta},  1  \right\} ^{1-\alpha}
	\right).
	\]

	\paragraph{Finding the optimal distribution.} If $z > \beta$, then $\frac{1 - z}{1 - \beta} < 1$ and the inner expression evaluates to $(\frac{z}{\beta})^\alpha$, which is maximized at $z = 1$. Similarly, when $z < \beta$, the inner expression evaluates to $(\frac{1 - z}{1 - \beta})^{1 - \alpha}$, which is maximized at $z = 0$. Therefore, we have
	\[
	\D^{\NW}(\alpha)
	= \min_{\beta \in [0, 1]}
	\, \max \left\{\left(\frac{1}{\beta}\right)^\alpha, \left(\frac{1}{1 - \beta}\right)^{1 - \alpha} \right\}.
	\]
	This is minimized when $\alpha \ln(\beta) = (1- \alpha) \ln(1 - \beta)$, and the unique minimizer of this expression is precisely the $\beta$ selected by $\NashTwo$. This proves that $\NashTwo$ always returns the distribution with the best possible distortion with respect to the Nash welfare on all profiles. Further, at this $\beta$, the distortion achieved is $\D_2^{\NW}(\NashTwo) = \max_{\alpha \in [0, 1]} \D^{\NW}(\alpha) = \sqrt{2}$, which is attained at $\alpha = \nicefrac{1}{2}$ (for which the optimal $\beta$ is also $\beta = \nicefrac{1}{2}$).
\end{proof}

\subsection{Proportional Fairness}
Given a preference profile $\profile$, let $\PFTwo$ be the voting rule that selects $\beta(\profile) = \frac{\sqrt{\alpha(\profile)}}{\sqrt{1 - \alpha(\profile)} + \sqrt{\alpha(\profile)}}$.

\begin{theorem}\label{thm:two-pf}
	For $m=2$ alternatives, and for any preference profile $\profile$, the voting rule $\PFTwo$ selects a distribution $\xx$ minimizing $\D^{\PF}(\xx,\profile)$.
	
	The rule is $(3/2)$-proportionally fair, which is the best possible among all voting rules. 
\end{theorem}
\begin{proof}
	Fix a preference profile $\profile$. 
	Let us write $\alpha = \alpha(\profile)$ and let $\D^{\PF}(\alpha) = \min_{\xx \in \Delta(A)} \D^{\PF}(\xx, \profile)$.
	From \Cref{lem:pf-approval-simplification}, we have 
	\begin{align*}
		\D^{\PF}(\alpha)
		&= \min_{\xx \in \Delta(\candids)} \max_{a \in\candids} \frac{1}{n} \sum_{i \in N} \frac{1}{\xx(h_i(a))}\\
		&= \min_{\beta \in [0, 1]} \max \left\{ \frac{\alpha }{\beta}  + (1- \alpha)\, ,\, \alpha + \frac{1- \alpha}{1 - \beta} \right\}.
	\end{align*}
	This expression is minimized when
	\[
	\frac{\alpha }{\beta}  + (1- \alpha) = \alpha + \frac{1- \alpha}{1 - \beta}
	\iff
	\alpha \cdot \frac{1 - \beta}{\beta} = (1 - \alpha) \cdot \frac{\beta}{1 - \beta}
	\iff
	\frac{\alpha}{1- \alpha} = \frac{\beta^2}{(1 - \beta)^2},
	\]
	which precisely yields the $\beta$ satisfying returned by $\PFTwo$. This shows that $\PFTwo$ always returns the distribution with the best possible proportional fairness on profile $\profile$. Further, at this value of $\beta$, the proportional fairness achieved is $\D^{\PF}_2(\PFTwo) = \max_{\alpha \in [0, 1]} \D^{\PF}(\alpha) = \nicefrac{3}{2}$, which is attained at $\alpha = \nicefrac{1}{2}$ (for which the optimal $\beta$ is also $\beta = \nicefrac{1}{2}$).
\end{proof}

\end{document}